\documentclass[journal]{IEEEtran}

\usepackage{cite}
\usepackage{amsmath,amssymb,amsfonts}
\usepackage{algorithmic}
\usepackage{graphicx}
\usepackage{textcomp}
\usepackage{multirow}
\usepackage{xcolor}
\usepackage{subfigure}
\usepackage{subcaption}

\graphicspath{ {./figure/} }
\ifCLASSINFOpdf
\else
\fi
\begin{document}
%
\title{PC-JND: Subjective Study and Dataset on Just Noticeable Difference for Point Clouds in 6DoF Virtual Reality\\}
%
%
%
\author{Chunling~Fan,
        Yun~Zhang,
        Dietmar~Saupe,
        Raouf Hamzaoui,
        and~Weisi~Lin

\thanks{Chunling~Fan is with the Shenzhen Polytechnic University, Shenzhen 518055, China (e-mail:fanchunling@szpu.edu.cn).}
\thanks{Yun~Zhang is with School of Electronics and Communication Engineering, Sun Yat-sen University, Shenzhen 510275, China (e-mail:zhangyun2@mail.sysu.edu.cn)}
\thanks{Dietmar Saupe is with University of Konstanz, D-78457 Konstanz, Germany (email:dietmar.saupe@uni-konstanz.de).}
\thanks{Raouf Hamzaoui is with the School of Engineering and Sustainable Development,
DeMonfort University, LE1 9BH Leicester, U.K. (e-mail: rhamzaoui@dmu.ac.uk).}
\thanks{Weisi Lin is with Nanyang Technological Unversity, Jurong West, Singapore (email:wslin@ntu.edu.sg).}
\thanks{Manuscript received April 19, 2005; revised August 26, 2015.}}

%
%

\markboth{Journal of \LaTeX\ Class Files,~Vol.~14, No.~8, August~2015}%
{Shell \MakeLowercase{\textit{et al.}}: Bare Demo of IEEEtran.cls for IEEE Journals}
%



\maketitle

\begin{abstract}
The Just Noticeable Difference (JND) accounts for the minimum distortion at which humans can perceive a difference between a pristine stimulus and its distorted version. The JND concept has been widely applied in visual signal processing tasks, including coding, transmission, rendering, and quality assessment, to optimize human-centric media experiences. A point cloud is a mainstream volumetric data representation consisting of both geometry information and attributes (e.g. color). Point clouds are used for advanced immersive 3D media such as Virtual Reality (VR). However, the JND characteristics of viewing point clouds in VR have not been explored before. In this paper, we study the point cloud-wise JND (PCJND) characteristics in a Six Degrees of Freedom (6DoF) VR environment using a head-mounted display. Our findings reveal that the texture PCJND of human eyes is smaller than the geometry PCJND for most point clouds. Furthermore, we identify a correlation between colorfulness and texture PCJND. However, there is no significant correlation between colorfulness and the geometry PCJND, nor between the number of points and neither the texture or geometry PCJND. To support future research in JND prediction and perception-driven signal processing, we introduce PC-JND, a novel point cloud-based JND dataset. This dataset will be made publicly available to facilitate advancements in perceptual optimization for immersive media.
\end{abstract}


%
\IEEEpeerreviewmaketitle

\section{Introduction}
%
%
%
%
\IEEEPARstart{T}{he} human visual system is a sophisticated neural network responsible for processing and interpreting visual stimuli. However, the human visual system exhibits inherent limitations in sensitivity and acuity due to the physiological limitations of the retinal photoreceptors. Specifically, it can only detect variations in visual signals that surpass a certain perceptual threshold~\cite{wu2019survey,LinWeisi}. Weber's law states that perceptual differences between stimuli must reach a fixed threshold of proportionality to produce distinguishable sensations. This fundamental psychophysical principle explains why humans can only detect image distortions exceeding a specific perceptual threshold, the Just Noticeable Difference (JND). The JND represents the visibility limit determined by both the inherent sensitivity of human vision and complex visual masking effects~\cite{Zhang2021Survey}. The JND characteristics provide valuable information for the development of mechanistic models of human vision and the advancement of research in visual psychology. These psychophysical findings have direct applications in perception-oriented image and video quality assessment, allowing the design of more accurate and human-visually aligned evaluation metrics~\cite{JNDIQA}, coding~\cite{JNDcoding,SURcoding}, enhancement~\cite{JNDenhancement}, watermarking~\cite{watermarking}, and other related applications.

Existing JND models can be categorized into four groups: pixel-domain JND models, subband-domain JND models, picture-wise JND (PWJND) models, and video-wise JND (VWJND) models. JND models in the pixel domain quantify detection thresholds at the individual pixel level by incorporating local spatial contextual information~\cite{WuEnhancedJND,ZENG}. They typically consider Luminance Adaptation (LA) and Contrast Masking (CM) effects. The LA effect indicates that the human visual system exhibits less sensitive to noise in extremely bright and dark backgrounds compared to gray backgrounds, following a non-linear response curve. The CM effect in images refers to the fact that the human visual system is less sensitive to noise in non-uniform regions due to the masking effect of high spatial frequency components, while remaining more acute in uniform regions. Pixel-domain JND models do not consider the global masking effects of the entire image or information in the frequency domain.

In subband-domain JND models, an image is first decomposed into frequency sub-band using a transform, such as the discrete cosine transform or discrete wavelet transform. The perceptual thresholds for the individual transform coefficients are then derived by considering the Contrast Sensitive Function (CSF) and masking effects~\cite{wang2020improved}. The spatial CSF describes the spatial frequency response of human visual system sensitivity, which acts as a band-pass filter for low spatial frequencies and a low-pass filter for high spatial frequencies~\cite{Zhang2021Survey}. However, the human visual system processes visual content as a whole, rather than analyzing isolated pixels or blocks. Complicated masking effects may lead to a larger JND threshold for an image or a video than the subband level JND thresholds because the human visual system is masked by other visual information.

PWJND and VWJND models have been developed to predict the JND of an image or video. 
Jin ~\textit{et al.}~\cite{MCL-JCI} found that the perceived distortion of JPEG-compressed images occurs in a stepwise manner, with five to seven perceptual levels, despite the presence of 100 quantization factor scales in JPEG. Once the PWJND of an image is estimated, it can be used as a guideline for image compression~\cite{liuPWJND}, reducing the bitrate while maintaining perceptual quality. In addition to the spatial masking effects considered in the PWJND, temporal masking effects were included in the building of the VWJND models\cite{zhangVWJND}. 
The human visual system is a binocular stereoscopic vision system that provides depth perception through parallax. JND models have also been developed for stereoscopic images~\cite{fan2021multimedia}. In recent years, several deep learning-based models have been proposed~\cite{fanSURnet,zhangVWJND,liuPWJND}. These data-driven models require large-scale datasets to improve prediction accuracy and avoid overfitting. However, building large JND datasets is extremely laborious and time-consuming. 

In summary, the existing work has mainly focused on 2D images, 2D videos, stereoscopic images, and panoramic images. This research has shown that the PWJND and VWJND thresholds depend on the complexity of the visual content.  
With the rapid development of multimedia technology and computer graphics, Virtual Reality (VR) has become increasingly popular due to its high level of immersion and Six Degrees of Freedom (6DoF) interaction. It has been widely used in training, industrial manufacturing, autonomous driving, education, healthcare, cultural communication, and other industries. Point clouds, which can represent real-world objects, are therefore widely used in VR. However, the JND characteristics of 2D images, 2D videos, stereoscopic images, and panoramic images are not applicable for point clouds.

Point clouds differ from 2D images and videos in three ways. First, point clouds typically consist of millions of points that contain geometry information (x, y, z) and attributes such as color (R, G, B), reflectance, transparency, and so on. Moreover, these points are unstructured, disordered, and irregular, whereas pixels in 2D images are regularly distributed. Second, the display device and interaction methods are different. 2D images and videos are viewed passively through 2D displays, while in VR applications, users wear a Head-Mounted Display (HMD) to view point clouds and can freely interact with point cloud models in 6DoF. In the VR environment, the JND characteristics of point clouds are more complex. Third, the data volume of point clouds is several times larger than that of images and videos, in similar scenarios. This large volume of data presents significant challenges to transmission and storage. Therefore, it is necessary and meaningful to study the JND characteristics of point clouds in VR. 

To address this problem, we conducted a subjective test to study the point cloud-wise JND (PCJND) using an HMD in a 6DoF VR environment. Our main contributions are as follows: First, we explore the PCJND characteristics of point clouds within a 6DoF VR environment using an HMD. To the best of our knowledge, this is the first work of its kind to study the PCJND. We develop a platform for subjects to locate PCJND thresholds. Subjects wear an HMD and can walk freely to view point cloud models in a room with controlled light as recommended in ITU-R BT.500-15~\cite{RecommendationITUR}. Second, we examine both the texture PCJND and geometry PCJND for point clouds. While texture and geometry distortions often occur simultaneously in practice and may influence each other perceptually, we address these distortions separately to maintain a more focused and manageable scope for this initial study of JND in point clouds. Finally, we create a point cloud-wise JND dataset, named PC-JND, which contains 34 high-quality reference point clouds along with distorted versions encoded using Video-based Point Cloud Compression (V-PCC) ~\cite{VPCC}. For each reference, we generate 51 distorted versions by varying geometry and texture QP values from 1 to 51. This dataset will be made publicly available.

The remainder of the paper is organized as follows. Section~\ref{Sec:proposedmethod} describes the subjective experiments and the construction of the proposed PC-JND dataset. Section~\ref{Sec:post-processing} presents data analysis and post-processing. Section~\ref{Sec:discussion} provides analysis and discussions. Section~\ref{Sec:concluision} gives the conclusions of this paper.

\section{Related Work}

In this section, we provide an overview of the existing PWJND/VWJND-based datasets and their subjective assessment methods.



Subjective tests are the most straightforward and reliable way to obtain JND samples and create benchmarks. 
In a typical subjective test, participants view pairs of images or videos, one original and one distorted, either simultaneously or sequentially, they judge whether the two stimuli appear the same or different. 
As presented in Table~\ref{tab:dataset}, the PWJND/VWJND characteristics of 2D images, videos, stereoscopic images, and panoramic images have been studied. As a result, several datasets have been created, such as image datasets MCL-JCI~\cite{MCL-JCI}, Shen~\textit{et al.}~\cite{shen2020jnd}, and KonJND-1k~\cite{KonJND-1k}, video datasets MCL-JCV~\cite{MCL-JCV}, VideoSet~\cite{VideoSet} and HD-JND~\cite{HD-VJND}, stereoscopic image datasets SIAT-JSSI~\cite{SIAT-JSSI} and SIAT-JASI~\cite{SIAT-JASI}, and panoramic image dataset JND-Pano~\cite{JND-Pano}. 

\begin{table*}[t]\caption{Summary of existing PWJND/VWJND based datasets.}\label{tab:dataset}
\tiny
\begin{center}
\resizebox{1\textwidth}{!}{
\begin{tabular}{c c c c c c c c c c}
\hline
datasets & Year & Stimuli type & Displayer & No. ref. & Dist. type & Dist. levels & No. rates/seq & Display way\\
\hline
MCL-JCI~\cite{MCL-JCI} & 2016 & image &2D monitor & 50 & JPEG & 100 &30 & side by side\\

Shen~\textit{et al.}~\cite{shen2020jnd} & 2020 & image & 2D monitor& 202 & H.266/VVC & 51  & 50 & side by side\\

KonJND-1K~\cite{KonJND-1k} & 2022 & image & 2D monitor& 1008 & JPEG/BPG & 100/51 & 42 & -\\

MCL-JCV~\cite{MCL-JCV} & 2016 & video & 2D monitor & 30 & H.264/AVC & 51 & 50 & one after another\\

Video-Set~\cite{VideoSet} & 2017 & video &  2D monitor & 220 & H.264/HEVC & 51 & 30 & one after another\\

HD-VJND~\cite{HD-VJND} & 2022 & video & 2D monitor & 180 & HEVC & CRF (dynamic) &20&one after another\\

SIAT-JSSI\cite{SIAT-JSSI} & 2019 & stereo image & 3D monitor & 10 & JPEG2000 H.265 Sym. & 300/51  & 36 & side by side \\

SIAT-JASI\cite{SIAT-JASI} & 2019 & stereo image & 3D monitor & 10 & JPEG2000 H.265  Asym. & 300/51 & 36 & side by side\\ 

JND-Pano~\cite{JND-Pano} & 2018 & panoramic image & HMD & 40 & JPEG & 100  &25 &one after another\\

PC-JND (proposed) & 2023 & point cloud & HMD & 34 &V-PCC & 51 & 33 & side by side\\
\hline
\end{tabular}
}
\end{center}
\vspace{-10pt}
\end{table*}

Jin~\textit{et al.}~\cite{MCL-JCI} conducted a laboratory study to explore the PWJND of 2D images and created a dataset called MCL-JCI. The dataset contains 50 reference images and their distorted versions using a JPEG encoder with quality factor ranging from 1 to 100, where 100 represents the best quality and 1 represents the worst quality. Images were displayed side by side on a 2D display in pairs, with each pair consisting of a reference image and a distorted version. Each pair was evaluated by 30 subjects. A binary search method was used to locate the PWJND threshold. They found that most of the test images had four to seven distinguishable distortion levels, the number of which was related to image content, texture, and semantics. They also derived a stair quality function from the collected PWJND samples. 

Shen~\textit{et al.}~\cite{shen2020jnd} determined the PWJND of 2D images for Versatile Video Coding (VVC) through a laboratory study. They collected 202 source images, which were coded with VVC where the Quantization Parameter (QP) ranged from 13 to 51. In the experiment, image pairs were displayed side by side according to the double-stimulus method. They found significant differences between the PWJND thresholds of different captured content.

Lin~\textit{et al.}~\cite{KonJND-1k} investigated the PWJND of 2D images through crowdsourcing and built a dataset called KonJND-1K. They collected 1008 reference images and considered both JPEG and BPG distortion types. They used slider adjustment to locate the PWJND, which reduces the test duration by 50\%. During the test, subjects adjusted the slider until they found the flickering point, the PWJND threshold. Crowdsourcing studies differ significantly from laboratory studies. In a laboratory setting, the environment is controlled and the process is closely supervised by the researchers. However, in crowdsourcing, testers may use different display devices, and experimental designers cannot supervise the subjects face-to-face. 

Fan~\textit{et al.}\cite{SIAT-JSSI,SIAT-JASI} explored PWJND for symmetric and asymmetric compressed stereo images through a laboratory study. They collected 10 source stereo images and encoded them using JPEG2000 and H.265 intra coding. They investigated PWJND$_{PRI}$ and PWJND$_{DRI}$ for symmetric and asymmetric stereo images respectively. The PWJND$_{PRI}$ was obtained using a pristine stereo image as the anchor, and the PWJND$_{DRI}$ was obtained by using a distorted stereoscopic image as the anchor. Stereoscopic image pairs were displayed side by side on a 3D monitor. Subjects rated in a controlled room wearing polarized glasses. Each stereo image was evaluated by 36 subjects. They found that PWJND depends on the image content and texture complexity. 

Liu~\textit{et al.}\cite{JND-Pano} studied the PWJND of panoramic images in a laboratory environment. They collected 40 reference panoramic images and 4000 distorted versions encoded using JPEG by changing the quality factor from 1 to 100. Unlike traditional test environments, subjects wore HMDs and sat in swivel chairs. Pairs of panoramic images were shown in random order. The study found that the PWJND threshold for panoramic images ranged from 30 to 80, depending on image content.  


Wang~\textit{et al.}~\cite{VideoSet} conducted a subjective test on a large-scale video dataset consisting of 220 source video clips at four resolutions. Each of the 880 video clips was encoded using the H.264 codec with QP values ranging from 1 to 51. To accommodate uncertain choices, a more robust binary search procedure was proposed. Instead of removing half of the original interval, only a quarter is removed. They measured the first three VWJNDs for all the video clips and converted the VWJND samples into an SUR curve. 

\section{Subject Experiment and Proposed PC-JND Dataset Construction} \label{Sec:proposedmethod}
The characteristics of point clouds differ significantly from traditional image and video. Therefore, we designed the subjective experiment accordingly. This section presents the reference creation, experimental settings, and evaluation methodology for the subjective test. The experiments were approved by the Institutional Review Board at Shenzhen Polytechnic University.

\begin{figure*}[htbp]
\centering

\begin{minipage}{0.99\textwidth}
\includegraphics[height=0.14\textwidth]{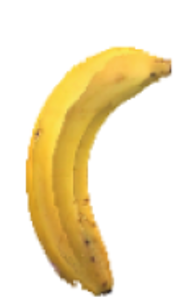}
\includegraphics[height=0.13\textwidth]{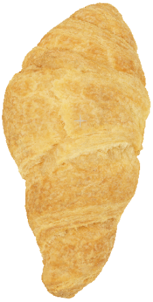}
\hspace{1pt}
\includegraphics[height=0.12\textwidth]{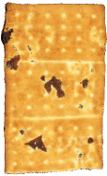}
\includegraphics[height=0.08\textwidth]{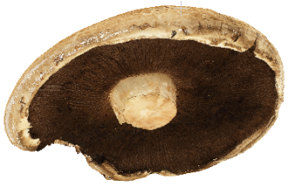}
\includegraphics[height=0.12\textwidth]{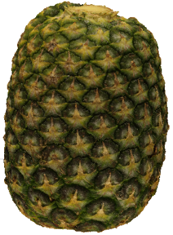}
\includegraphics[height=0.08\textwidth]{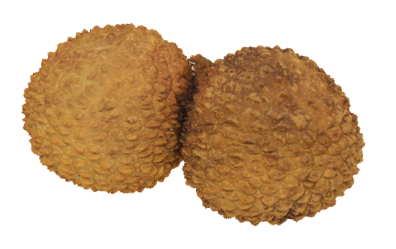}
\includegraphics[height=0.12\textwidth]{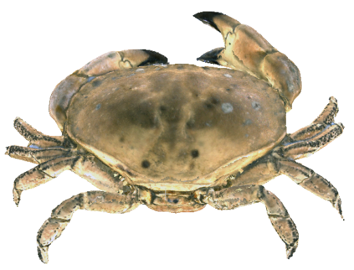}
\includegraphics[height=0.12\textwidth]{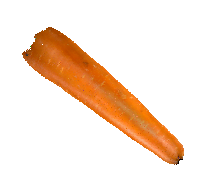}
\centerline{(a) Foods}
\end{minipage}
\vspace{5pt}

\begin{minipage}{0.38\textwidth}
\includegraphics[height=0.41\textwidth]{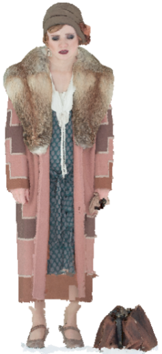}
\includegraphics[height=0.41\textwidth]{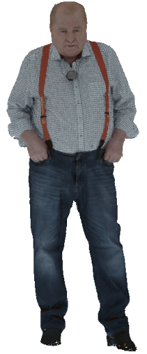}
\includegraphics[height=0.41\textwidth]{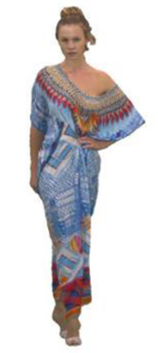}
\includegraphics[height=0.41\textwidth]{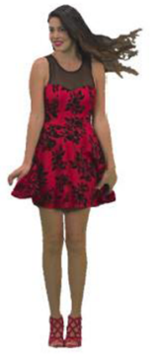}
\centerline{(b) People}
\end{minipage}
\hspace{2pt}
\begin{minipage}{0.6\textwidth}
\includegraphics[height=0.19\textwidth]{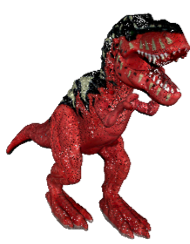}
\hspace{1pt}
\includegraphics[height=0.18\textwidth]{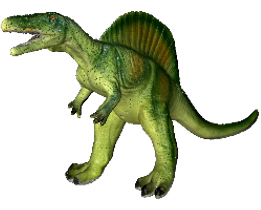}
\includegraphics[height=0.18\textwidth]{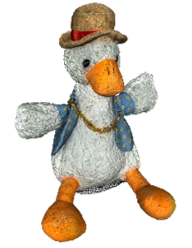}
\includegraphics[height=0.18\textwidth]{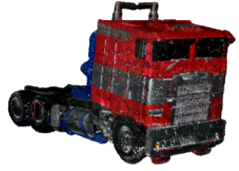}
\includegraphics[height=0.18\textwidth]{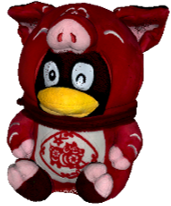}
\centerline{(c) Toys}
\end{minipage}
\vspace{5pt}

\begin{minipage}{0.64\textwidth}
\includegraphics[height=0.22\textwidth]{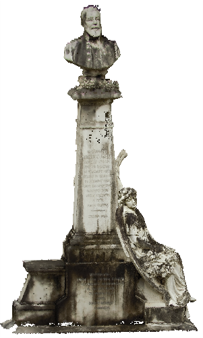}
\includegraphics[height=0.21\textwidth]{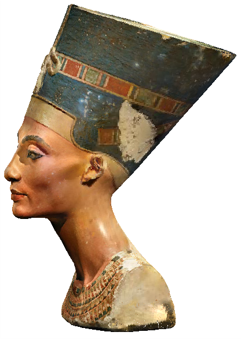}
\includegraphics[height=0.22\textwidth]{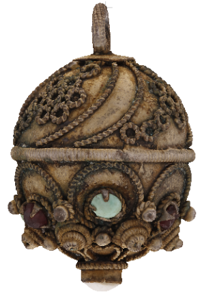}
\includegraphics[height=0.22\textwidth]{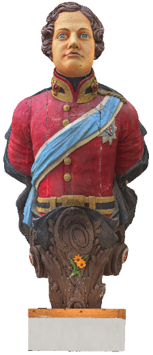}
\includegraphics[height=0.21\textwidth]{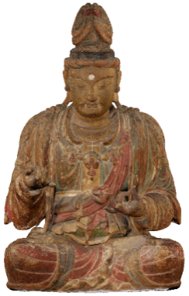}
\includegraphics[height=0.21\textwidth]{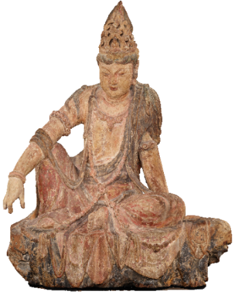}
\centerline{(d) Sculptures}
\end{minipage}
\begin{minipage}{0.34\textwidth}
\hspace{5pt}
\includegraphics[height=0.32\textwidth]{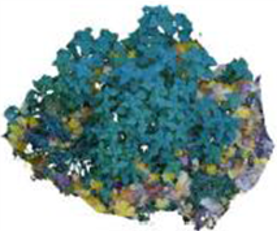}
\includegraphics[height=0.38\textwidth]{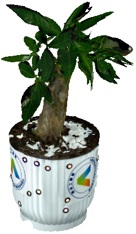}
\includegraphics[height=0.38\textwidth]{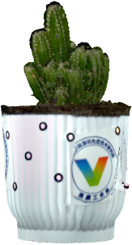}
\centerline{(e) Plants}
\end{minipage}
\vspace{5pt}

\begin{minipage}{0.48\textwidth}
\includegraphics[height=0.22\textwidth]{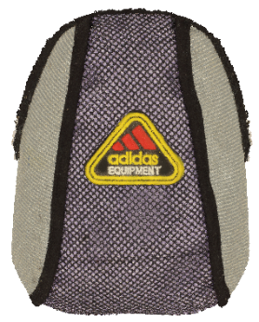}
\includegraphics[height=0.22\textwidth]{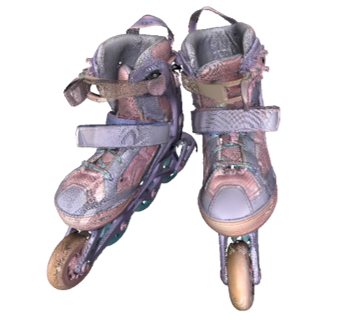}
\includegraphics[height=0.13\textwidth]{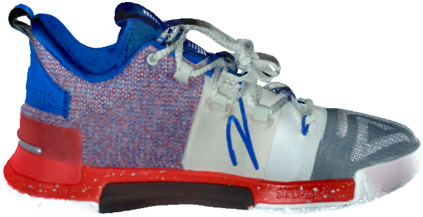}
\includegraphics[height=0.13\textwidth]{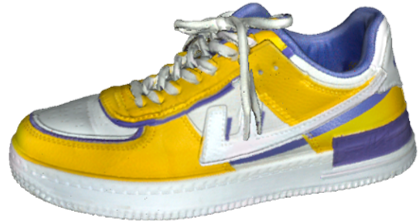}
\centerline{(f) Clothing}
\end{minipage}
\hspace{3pt}
\begin{minipage}{0.48\textwidth}
\includegraphics[height=0.2\textwidth]{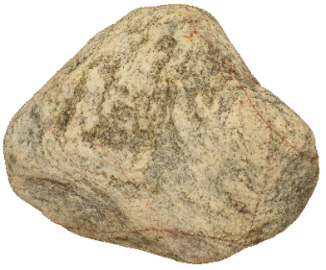}
\includegraphics[height=0.22\textwidth]{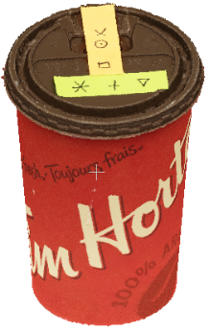}
\includegraphics[height=0.22\textwidth]{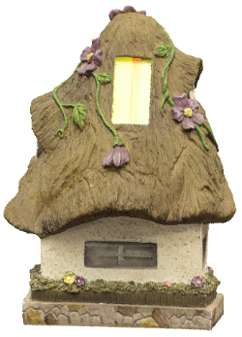}
\includegraphics[height=0.2\textwidth]{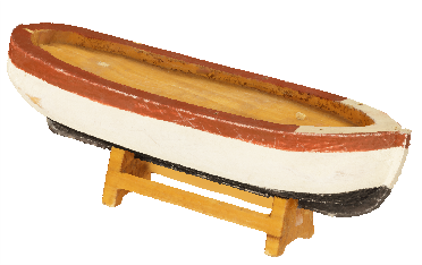}
\centerline{(g) Others}
\end{minipage}
\caption{Snapshots of the point clouds used in our subjective test.}\label{fig:snapshots}
\end{figure*}

\begin{table*}[t]\caption{Summary of point clouds.}\label{tab:reference}
\begin{center}
\resizebox{0.8\textwidth}{!}{
\begin{tabular}{c c c r c c c c}
\hline
Category & Reference & Source & No. Points & (xmax,ymax,zmax) & geometry Precision \\
\hline

\multirow{8}{*}{Foods} 
& bananamesh & Sketchfab & 202,770 & (1023,684,369)& 10bits\\
& croissant & WPC & 827,172 & (722,1023,551)& 10bits \\
& biscuits & WPC & 894,345 & (638,1023,139)& 10bits \\
& mushroom & WPC & 1,118,520 & (1023,431,923)& 10bits\\
& pineapple & WPC & 1,583,654 & (733,1023,766)& 10bits\\
& litchi & WPC &1,039,942 & (1023,522,563)& 10bits\\
& carcinus & AVS & 1,500,549 & (1023,740,402)& 10bits\\
& carrot & AVS & 84,966 & (1023,451,454)& 10bits\\
\hline

\multirow{4}{*}{People} 
& the20smaria & MPEG & 894,364 & (457,1023,365) & 10bits\\
& ulliwegner & MPEG & 702,785 & (386,1023,267)& 10bits\\
& longdress & MPEG & 797,178 & (358,1023,392)& 10bits\\
& redandblack & MPEG & 729,133 & (412,1023,243)& 10bits\\
\hline

\multirow{5}{*}{Toys}
& t-rex1 & AVS &199,367 & (1023,807,792)& 10bits\\
& spinosaurus& AVS &237,351 &(1023,657,633)& 10bits \\
& duckdoll & AVS  & 694,349 &(778,1023,727)& 10bits \\
& optimus\_Prime & AVS & 202,070 & (941,616,1023)& 10bits\\
& dragPenguin & AVS &1,125,441 & (898,1023,765)& 10bits\\
\hline

\multirow{6}{*}{Sculptures}
& angelseated & Sketchfab & 770,184& (595,1023,374)& 10bits\\
& bust & AVS & 1,362,119 & (503,1023,704) & 10bits\\
& knob & AVS & 2,722,038& (669,1023,673)& 10bits\\
& prince & AVS & 1,659,558 & (428,1023,394)& 10bits\\
& guanyin & AVS &2,554,381 & (660,1023,587)& 10bits\\
& guanyin1 & AVS & 2,288,431 &(827,1023,481)& 10bits\\
\hline

\multirow{3}{*}{Plants}
& new\_grass & Sketchfab & 466,884 & (1023,454,861)& 10bits\\
& potted-plant01 & AVS & 273,605 &(660,1023,513)& 10bits\\
& potted-plant03 & AVS & 214,137 & (563,1023,604)& 10bits\\
\hline

\multirow{4}{*}{Clothing}
& bag & WPC & 1,235,103 & (813,1023,581)& 10bits\\
& rollerskating & AVS & 929,164 & (852,901,1023) & 10bits\\
& weaving-shoe & AVS & 788,512 &(1023,613,581) & 10bits\\
& leap & AVS & 990,722 & (1023,778,661)& 10bits \\
\hline

\multirow{3}{*}{Others}
& stone & WPC &1,086,453 & (1023,834,599)& 10bits\\
& coffee\_cup & WPC & 1,264,341 & (644,1023,961) & 10bits\\
& house & WPC & 1,527,166 & (742,1023,692)& 10bits\\
& ship & WPC & 677,989 & (1023,397,358)& 10bits\\
\hline
\end{tabular}
}
\end{center}
\vspace{-10pt}
\end{table*}

\subsection{Reference and Distorted point cloud Creation}\label{Sec:referenceCreation}
Studying the JND of point clouds requires the capture and selection of diverse reference point clouds. Three key aspects were considered: semantic content, visual quality, and content diversity. First, to collect point clouds with diverse semantic content and visual quality, we selected 34 high-quality point clouds as references from AVS, WPC, WPC2.0, MPEG, JPEG, and Sketchfab, as shown in Figure~\ref{fig:snapshots}. Seven key semantic types of point clouds were included: food, people, toys, sculptures, plants, clothing, and others, each with various geometric structures and textures. Table~\ref{tab:reference} presents a summary of all point clouds, including the source, number of points, bounding box and geometry precision. The number of points ranges from 84,966 to 2,722,038. All bounding boxes of the point clouds are of the form [0,xmax]$\times$ [0,ymax]$\times$[0,zmax]. Moreover, the geometry precision is fixed as a 10-bit depth.

\begin{figure*}[htbp]
\centering
\includegraphics[width=0.7\textwidth]{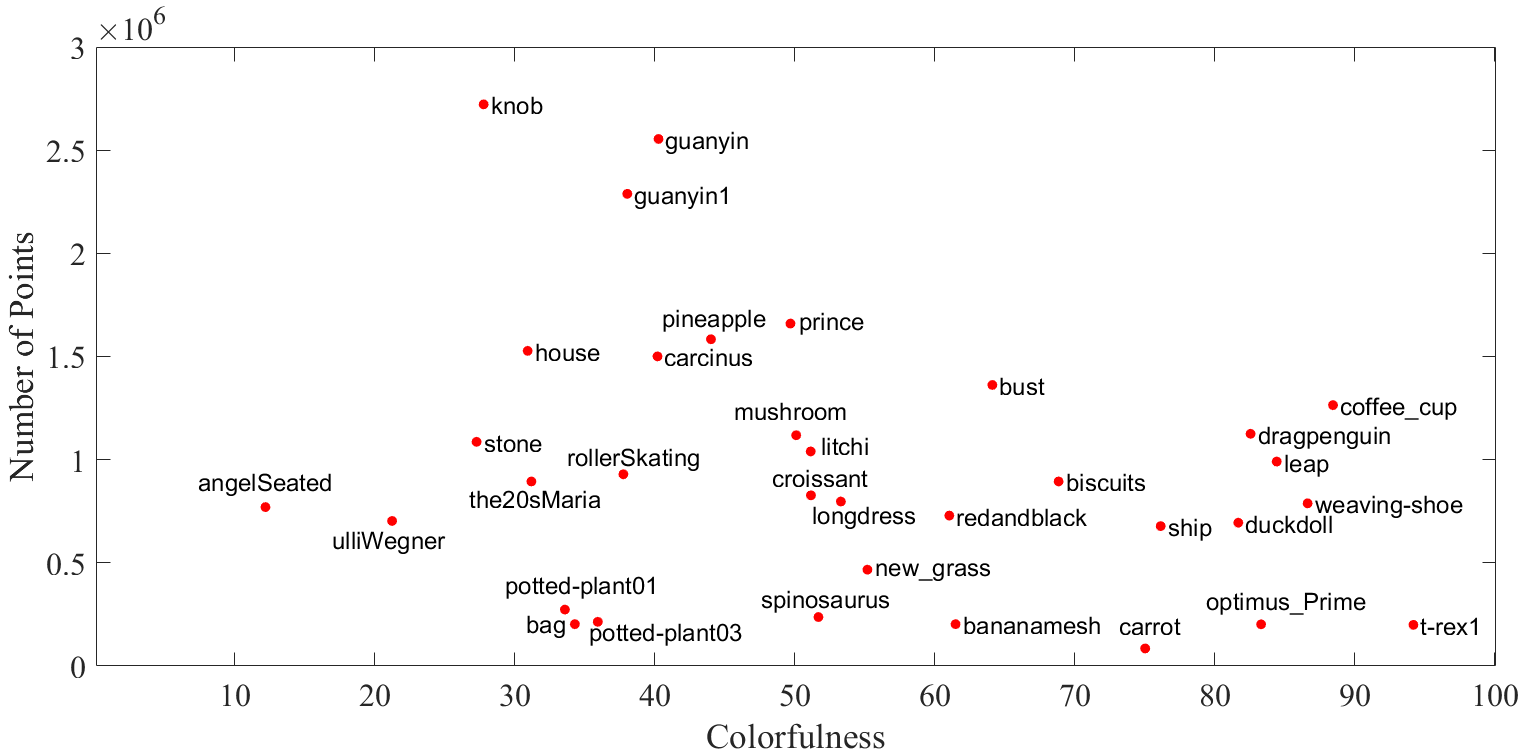}
\caption{Distribution of colorfulness and the number of points in the point clouds.}\label{fig:SICF}
\end{figure*}
In addition, the content diversity of the selected point clouds is analyzed. We used the number of points and the colorfulness~\cite{SI-CF} of the texture to measure the diversity of point clouds, where colorfulness denotes the perceptual variety and color intensity of the image. Figure~\ref{fig:SICF} shows the distributions of colorfulness and the number of points. It is observed that colorfulness and the number of points of the point clouds are widely spread, which indicates that the selected point clouds are diverse in content.

\begin{figure}[htbp]
\centering
\includegraphics[width=0.45\textwidth]{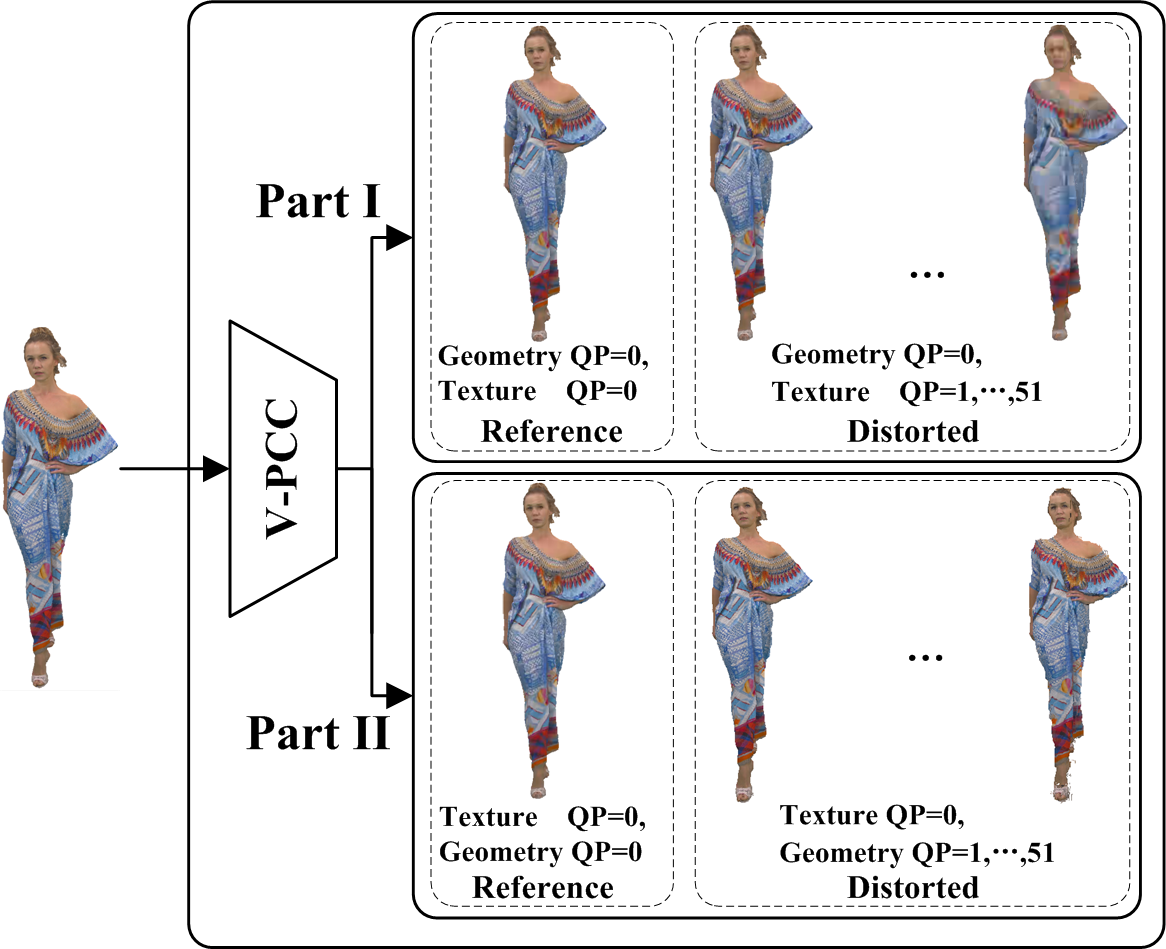}
\caption{The reference was encoded using texture and geometry QP=0, the distorted versions were encoded by changing texture QP in Part I, geometry QP in Part II.}\label{fig:preprocessing}
\end{figure}

The selected point clouds have different textures, scales, and bounding boxes. To facilitate subsequent operations, we applied a series of pre-processing steps to the selected point clouds, as shown in Figure~\ref{fig:preprocessing}. Specifically, to provide a realistic scenario the point clouds were rotated, scaled and translated to ensure that each had a tight bounding box with a maximum side length of 1023. In our subjective test, we aimed to determine the PCJND using V-PCC. Since V-PCC cannot process decimal values, the point coordinates were rounded to integers within the range 0,…,1023 (10-bit precision). For the transformations, we used the CloudCompare software\footnote{https://www.CloudCompare.org/}. For details of the resulting dataset, see Table~\ref{tab:reference}.

In this study, we aimed to determine the texture PCJND and the geometry PCJND. These metrics define the minimum distortion thresholds for texture and geometry, respectively, at which viewers can perceive differences between compressed and pristine point clouds. MPEG has developed two point cloud compression standards: V-PCC and Geometry-based point cloud Compression (G-PCC)~\cite{VPCCandGPCC}. V-PCC and G-PCC introduce different types of distortions. G-PCC primarily reduces the number of points, whereas V-PCC projects 3D point clouds onto 2D geometry and texture videos, which are then compressed using video encoders like HEVC and VVC, potentially introducing both geometry and texture distortions. The study in~\cite{Perry} found that V-PCC outperforms G-PCC in compressing dense colored point clouds. Our subjective test focuses on analyzing V-PCC distortions in texture and geometry and consists of two parts.
Part I involved compressing the texture videos using HEVC with QPs ranging from 0 to 51, while leaving the geometry videos uncompressed. The goal was to determine the texture PCJND in V-PCC. Texture distortions typically involve changes in the color attributes of the point cloud, resulting in luminance/chroma changes, spatial frequency variations, blur, and blocking artifacts. Part II involved compressing the geometry videos using HEVC with QPs ranging from 0 to 51, while keeping the texture videos uncompressed. The goal was to determine the PCJND in geometry compression with V-PCC. Geometry distortions in point clouds involve changes in the 3D positions and number of points, among other factors. A high QP value indicates low quality, with 51 representing the worst and 0 the best quality.

\begin{figure*}[htbp]
\centering
\includegraphics[width=0.9\textwidth]{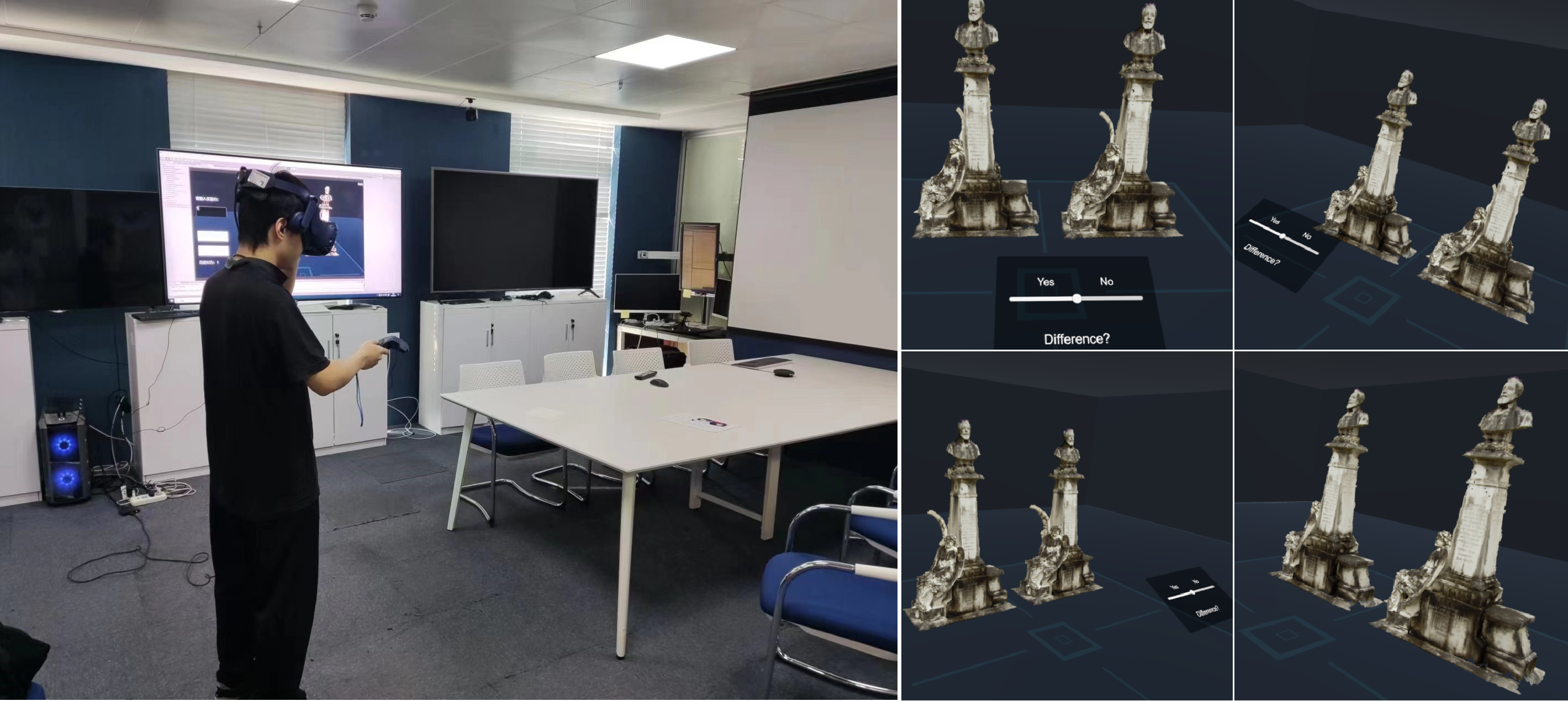}
\caption{6DoF VR Environment and rendering views.}\label{fig:environment}
\end{figure*}

\begin{figure*}[h]
\centering
\includegraphics[width=0.9\textwidth]{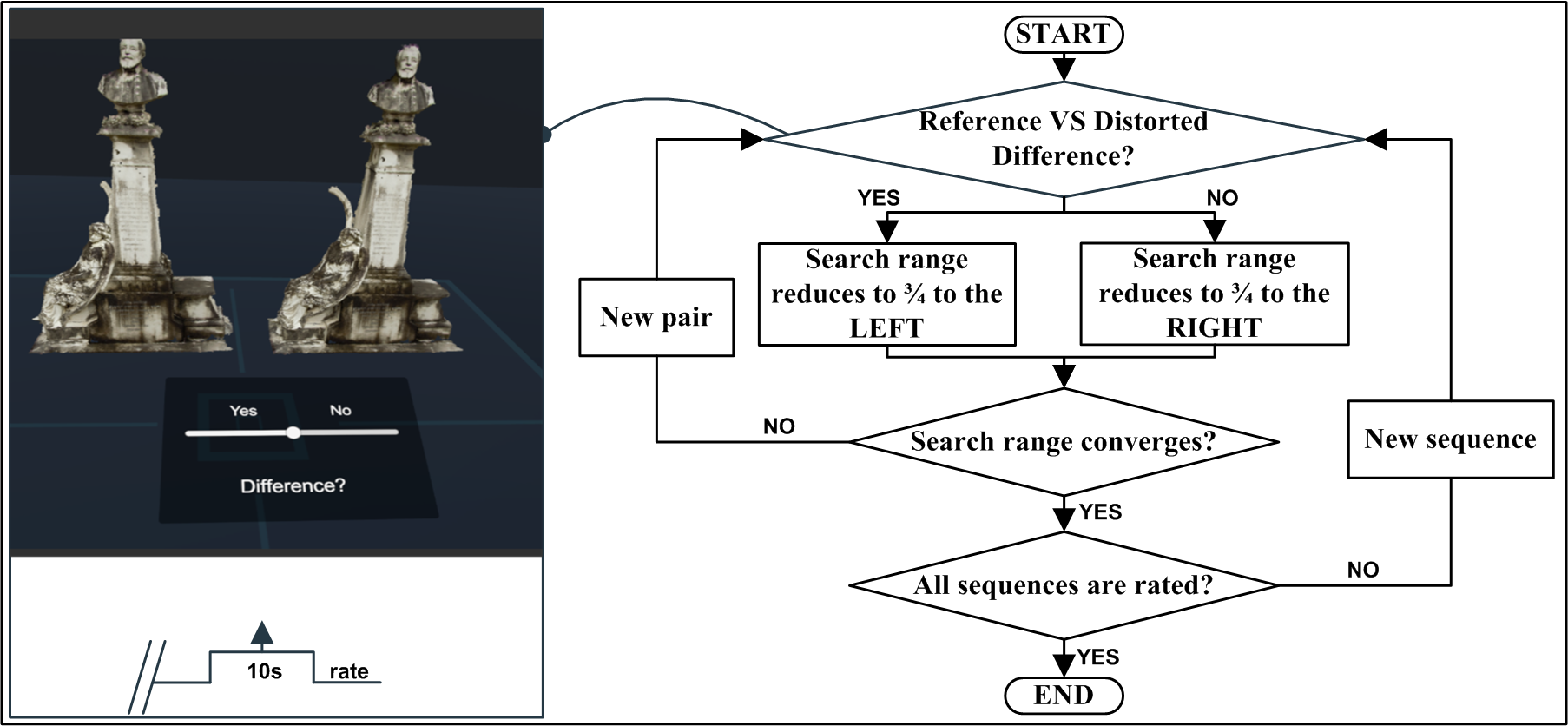}
\caption{Evaluation flowchart.}\label{fig:evaluation}
\end{figure*}

\subsection{VR Environmental Settings and Subjects}

To conduct the subjective experiment, we used the HTC Vive HMD as the display. THC Vive supports 6DoF viewing interactions and provides an immersive VR experience. Moreover, two handheld controllers were used to collect subjective scores. The positions and angles of the two controllers and VR HMD were tracked using two Lighthouse locators. The monocular resolution of the HMD was 1080$\times$1200 pixels with a refresh rate of 90 Hz. The combined resolution of the display was 2160$\times$1200 pixels with a field of view of 110 degrees. Based on the recommendations of ITU-R BT.500-15~\cite{RecommendationITUR} and recent studies~\cite{wusubjective,guo2016subjective}, the virtual environment in VR was set as an empty room with dark gray floor and walls with smooth texture to minimize the visual interference with the point cloud models from the background scenes. At the same time, 6500K virtual lights were placed on the ceiling of the virtual room to simulate office lighting. As shown in Figure~\ref{fig:environment}, our subjective test was conducted in a controlled room that was large enough to allow the subjects to walk and change view freely. The point clouds were placed in the center of the virtual room, initially about 2 meters away from the viewer. In the virtual environment, subjects could move closer to or further from the point cloud objects for viewing. Moreover, subjects could freely view the objects from all angles, front, side, and back, both externally and internally. Figure~\ref{fig:environment} illustrates four rendered views of an example point cloud pair, consisting of a pristine and a distorted version.

In total, 68 non-expert subjects, who are college or university students aged 18 to 26, participated in our subjective test. All subjects had normal or corrected-to-normal vision. They all passed visual acuity and color vision tests before the subjective test.  To familiarize each subject with the HTC Vive, we conducted a training session in which they viewed two sets of point clouds, distinct from those used in the test (the point clouds used in the test are listed in Table\textbf{\ref{tab:dataset}}). During training, subjects were informed of the purpose of the experiment, how to operate the HTC Vive, and some precautions. For example, during the test, subjects were allowed to walk around the point clouds, but were advised not to get too close. This is because empty holes or void spaces between points become visible when viewed from a very close distance. In addition, subjects were allowed to stop the experiment at any time if they felt dizzy or unwell. Those who passed the training phase were allowed to participate in the formal test. Each part of the test was divided into two sessions, each including 17 point clouds. In each session, subjects could freely view the point cloud pairs from any position and angle until they were ready to make a decision. Consequently, the duration varied from person to person. Overall, each part of the subjective test took approximately one hour for each subject. To avoid visual fatigue, subjects were asked to rest for five to ten minutes after 30 minutes of testing or whenever they felt tired. 

\subsection{Evaluation Methodology}

As shown in Figure~\ref{fig:evaluation}, similar to the Double-Stimulus Impairment Scale (DSIS)~\cite{RecommendationITUR,comparisonVR} method, a pair of point clouds were displayed side by side, one for the reference and the other for the distortion. Subjects were asked if there was a difference between them. They were instructed to make a ``YES'' or ``NO'' choice within about ten seconds, as recommended in~\cite{RecommendationITUR}. When the observer moves the slider of the interface to the left, a ``YES'' response is recorded (there is noticeable distortion), while moving it to the right records a ``NO'' response.

The PCJND corresponds to the minimum level of distortion at which humans can perceive a difference between the original and distorted point clouds. For V-PCC, the aim is to determine an integer QP value in the range [1, 51]. A straightforward strategy is to compare the reference point cloud with all distorted versions, which requires 51 comparisons. However, this strategy is too time-consuming. To reduce the number of comparisons, a binary search that halves the search range with each step was used. Fig.~\ref{fig:evaluation} shows the flowchart of the binary search for PCJND. For example, suppose the initial search range was [1, 51], and the reference was compared with distortion level 26. If the subject chose 'NO', the search range would be reduced to [26, 51]. However, if the subject made an incorrect choice by mistake, it could affect the search result and the final PCJND value. To address this problem, a relaxed binary search~\cite{VideoSet} was used. The search range was narrowed to three quarters instead of half of the original search range. Thirty four point clouds in each part were randomly displayed in sequence. Sixty eight subjects were involved in the test, and 33 geometry PCJND samples and 35 texture PCJND samples for each reference were collected. 

\section{Data Processing} \label{Sec:post-processing}

In subjective tests, outliers may occur due to fatigue, operating errors, etc. To ensure the credibility and reliability of the results, we detected the outliers of subjects and samples in two steps. 

\begin{figure*}[t]
\centering
\begin{minipage}{0.58\textwidth}
\includegraphics[width=0.9\textwidth]{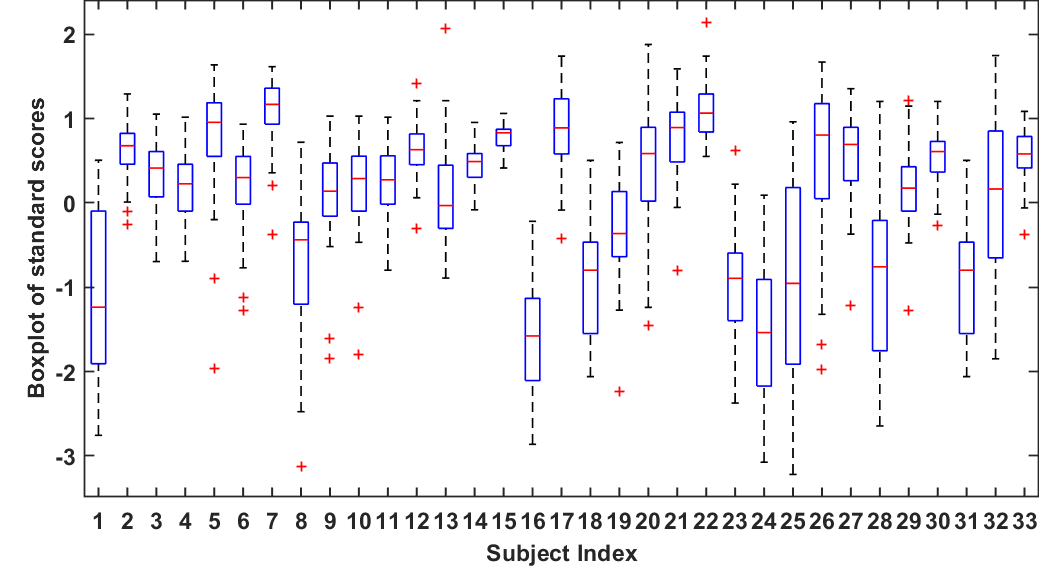}
\centerline{(a)}
\end{minipage}
\begin{minipage}{0.41\textwidth}
\hspace{15pt}
\includegraphics[width=0.9\textwidth]{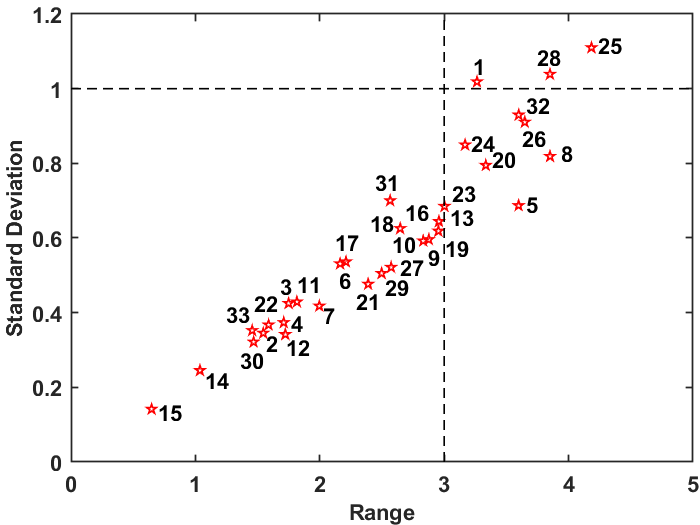}
\centerline{(b)}
\end{minipage}
\centering
\begin{minipage}{0.58\textwidth}
\includegraphics[width=0.9\textwidth]{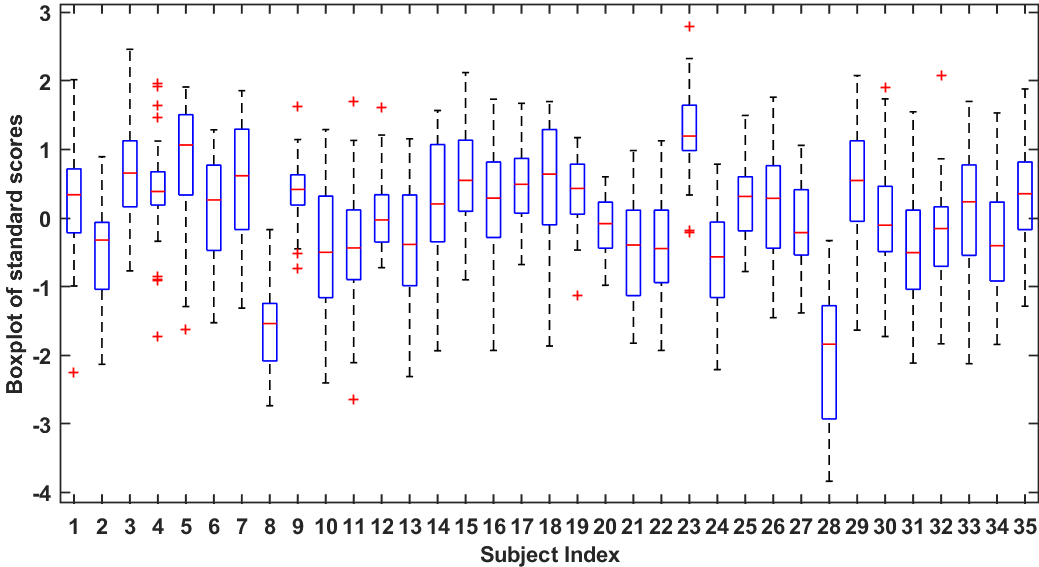}
\centerline{(c)}
\end{minipage}
\begin{minipage}{0.41\textwidth}
\hspace{15pt}
\includegraphics[width=0.9\textwidth]{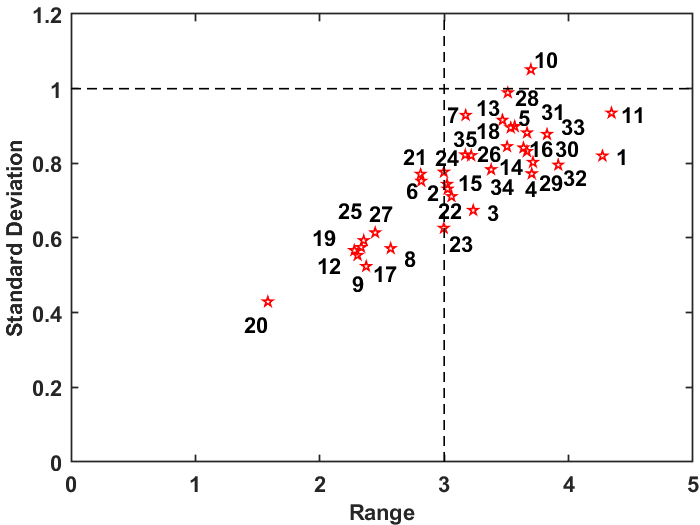}
\centerline{(d)}
\end{minipage}
\caption{Boxplot and range vs. standard deviation of the standard scores. (a) boxplot for texture PCJND samples. (b) range vs. standard deviation for texture PCJND samples. (c) boxplot for geometry PCJND samples. (d) range vs. standard deviation for geometry PCJND samples.}\label{fig:boxplot}
\end{figure*}

First, outlier subjects were detected and their PCJND samples were removed. Let $S^m_n$ be the PCJND sample from subject $m$ on reference $n$, where $m=1,...,M$ and $n=1,...,N$. Moreover, PCJND value is standardized. The standardized score PCJND was derived as
\begin{equation}
Z^m_n=(S^m_n-\mu_n)/\sigma_n,   
\end{equation}
where $\mu_n$ and $\sigma_n$ are the mean and standard deviation of PCJND of reference $n$. 
Figures~\ref{fig:boxplot}(a) and (c) present the boxplots of the standardized scores for the 33 subjects with minimum and maximum values of -3.2205 and 2.1407 in part I and 35 subjects with minimum and maximum values of -3.8409 and 2.7913 in part II. 

For each subject $m$, we calculated the range $r_m$ and standard deviation $\sigma_m$ from his/her standardized scores for all point clouds, where 
\begin{equation}
r_m=\mathop{\max}_{n=1,...,N}(Z^m_n)-\mathop{\min}_{n=1,...,N}(Z^m_n), m=1,...,M 
\end{equation}


Inspired by~\cite{VideoSet}, a subject will be detected as an outlier if both of its range $r_m$ and standard deviation $\sigma_m$ are large. To be specific, the subject is outlier if $r_m>3$ and $\sigma_m>1$. As shown in Figures~\ref{fig:boxplot}(b) and (d), three subjects (subject 1, 25, and 28) in part I and one subject (subject 10) in part II were detected as outlier subjects. Therefore, the PCJND samples obtained from these subjects were removed.

Second, after eliminating outlier subjects, there may still exist outliers in the remaining samples. We used Grubbs' test~\cite{grubbs} to detect and remove outlier samples following the same procedure as in~\cite{VideoSet} and~\cite{SIAT-JASI}. One outlier sample was removed at a time, and the procedure continued until all samples passed the Grubbs' test. For reference $n$, we get PCJND samples $S^m_n, m=1,...M$. In Grubbs' test, $G_n$ for reference $n$ was calculated as
\begin{equation}
G_n=\frac{\max_{m=1,...,M}{|S^m_n-\mu_n|} }{\sigma_n}.
\end{equation}
The Grubbs' critical value table $G(\alpha,M)$ can be found in \cite{grubbs}, where $\alpha$ denotes the significance level and $M$ denotes the sample size. Here, the sample size is around 30, and the significance level was set as $\alpha$ = 0.05. If $G_n>G(\alpha,M)$, then a PCJND sample is identified as an outlier. In part I, seven samples from 1020 (0.69\%) were detected and removed. In part II, eight samples from 1156 (0.69\%) were removed. Therefore, most of the samples were normal samples. 
 
As in~\cite{SIAT-JASI}, we applied the Jarque-Bera normality test~\cite{JBtest} to the collected JND samples. 
Most references passed the normality test. Specifically, in Part I, 32 references (94.12\%) passed the normality test, two references ``carrot" and ``potted-plant01" did not. In part II, we found that 28 references (85.29\%) passed the normality test, five references ``bananamesh", ``carcinus", ``longdress", ``T-rex1", and ``dragpenguin" did not pass. Overall, most of the PCJND samples passed the normality test indicating that fitting the SUR curve using a normal distribution is accurate and reasonable. After data processing, the remaining PCJND labels from the subjective study were considered valid.


\begin{figure}
    \centering
    \includegraphics[width=0.4\textwidth]{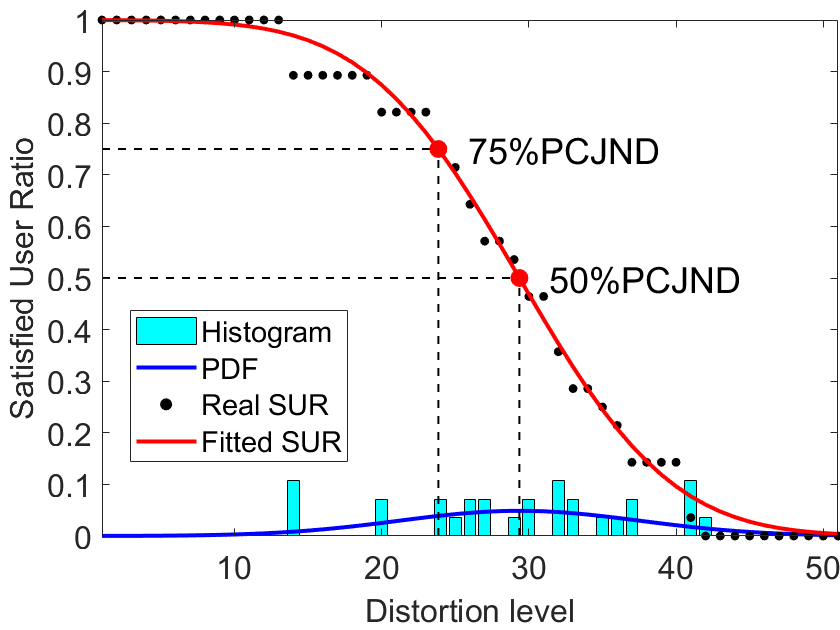}
    \caption{SUR curve, 75\% and 50\% PCJNDs of point cloud $bananamesh$.}
    \label{fig:example}
\end{figure}
\section{Analysis and Discussions}\label{Sec:discussion}

\paragraph{Satisfied User Ratio and PCJND} 
If a subject cannot perceive a difference between a reference point cloud and its distorted version, the distorted version is considered perceptually satisfactory. The SUR is the proportion of subjects who find the distorted version perceptually satisfactory. Let $r$ be a reference point cloud and $d_i, i=1,...,51$, be a sequence of texture or geometry distorted versions obtained with V-PCC encoding using QP=1,...,51. For a given distorted version $d_i$ and a population of $M$ subjects, the SUR is estimated as 
\begin{equation}
SUR(i)=1-\frac{1}{M}\sum_{m=1}^{M}\Phi_m(d_i),
\end{equation}
\noindent where $\Phi_m(d_i)=0$ if subject $m$ is satisfied with the quality of $d_i$ and 1, otherwise. 

\begin{figure*}[t]
\centering
\begin{minipage}{1\textwidth}
\centering
\includegraphics[width=0.3\textwidth]{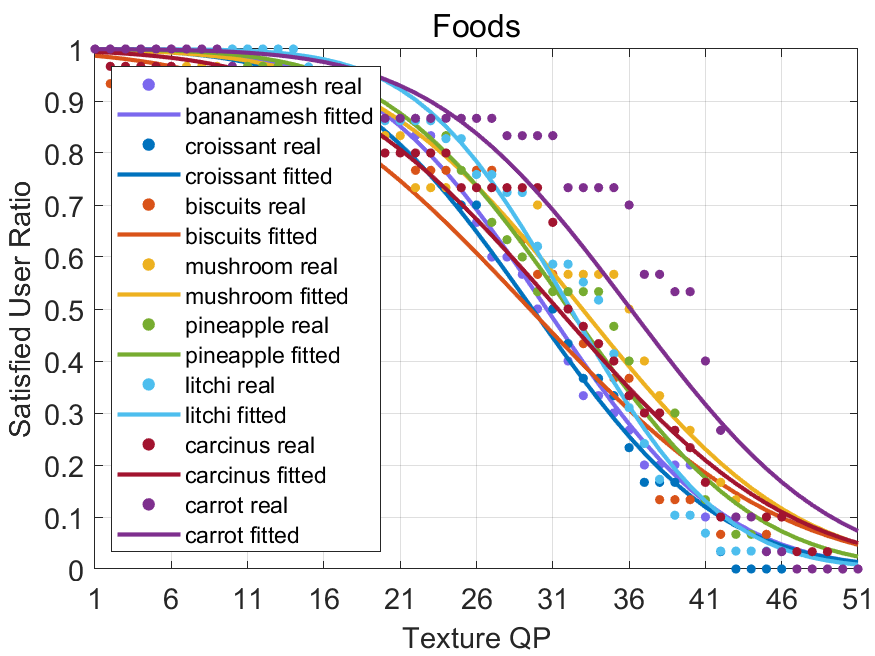}
\includegraphics[width=0.3\textwidth]{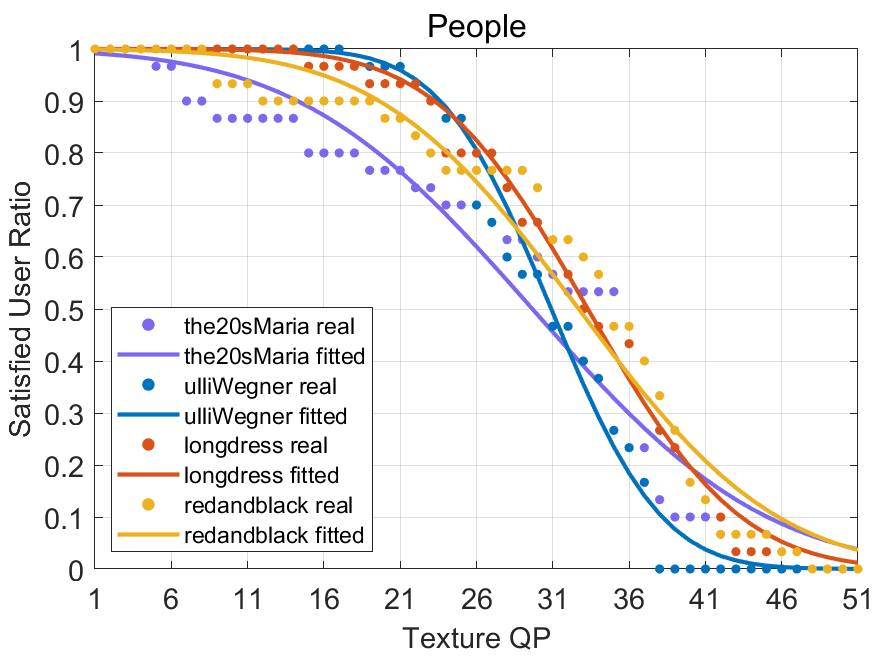}
\includegraphics[width=0.3\textwidth]{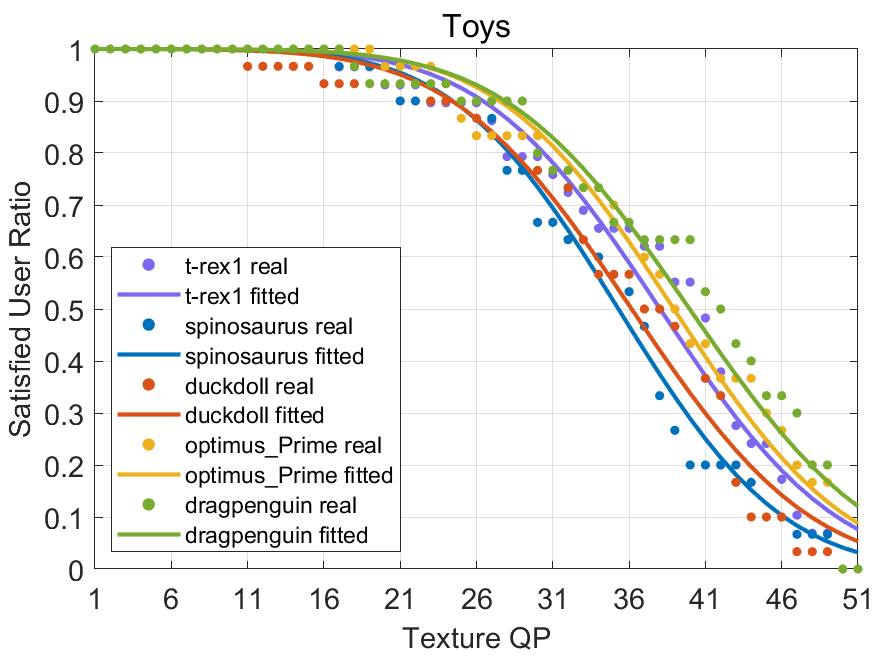}
\includegraphics[width=0.3\textwidth]{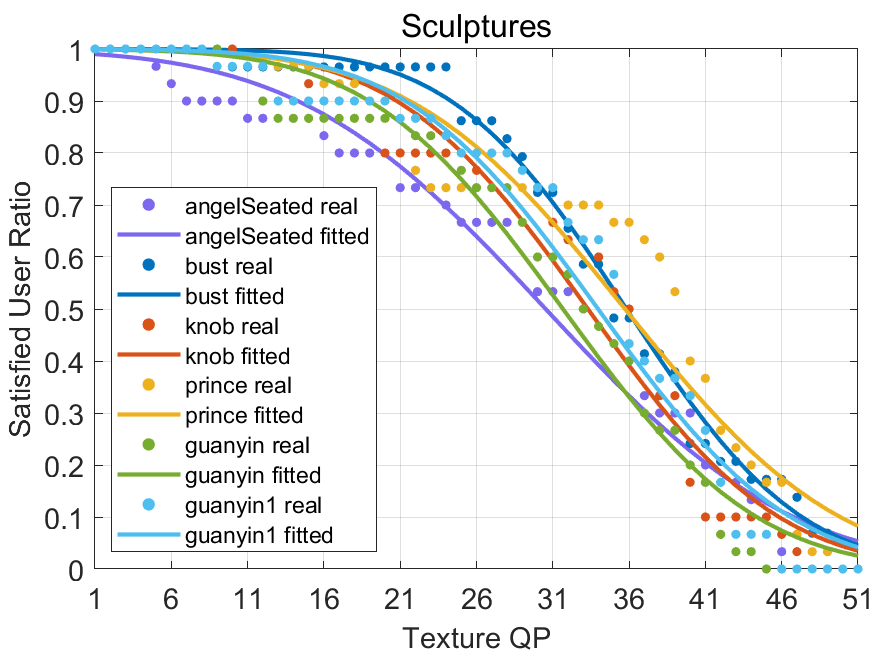}
\includegraphics[width=0.3\textwidth]{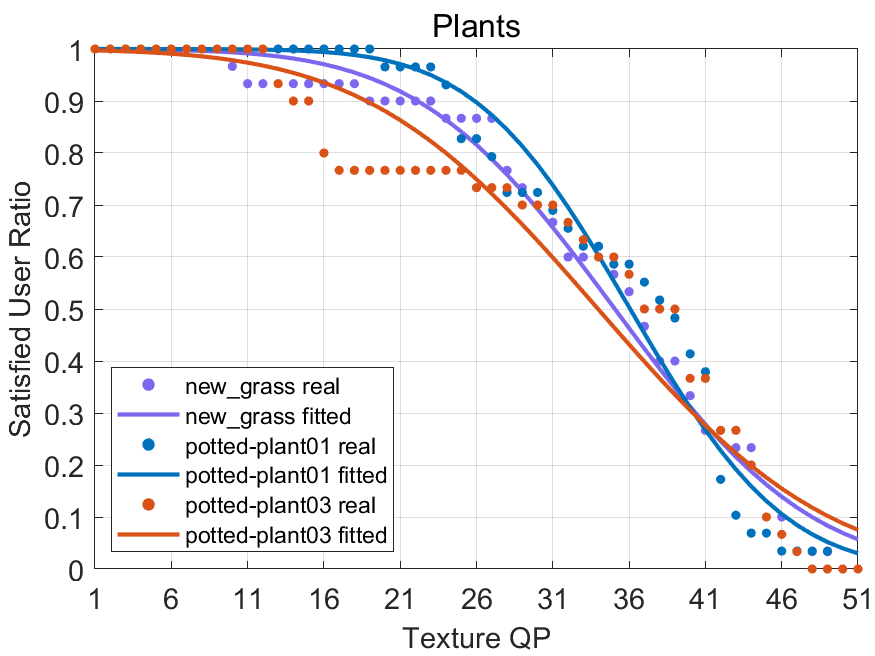}
\includegraphics[width=0.3\textwidth]{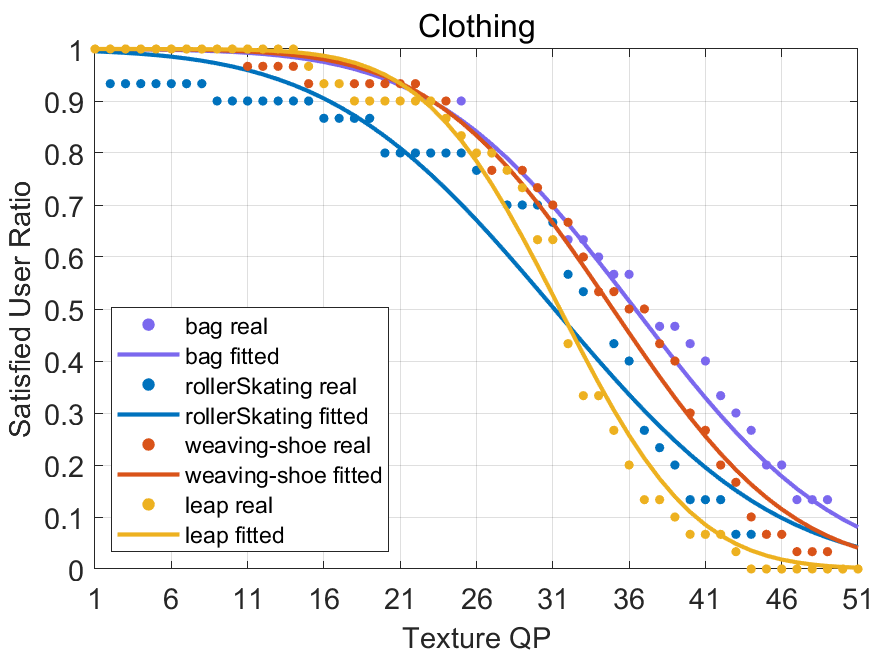}
\centerline{(a)}
\end{minipage}
\begin{minipage}{1\textwidth}
\centering
\includegraphics[width=0.3\textwidth]{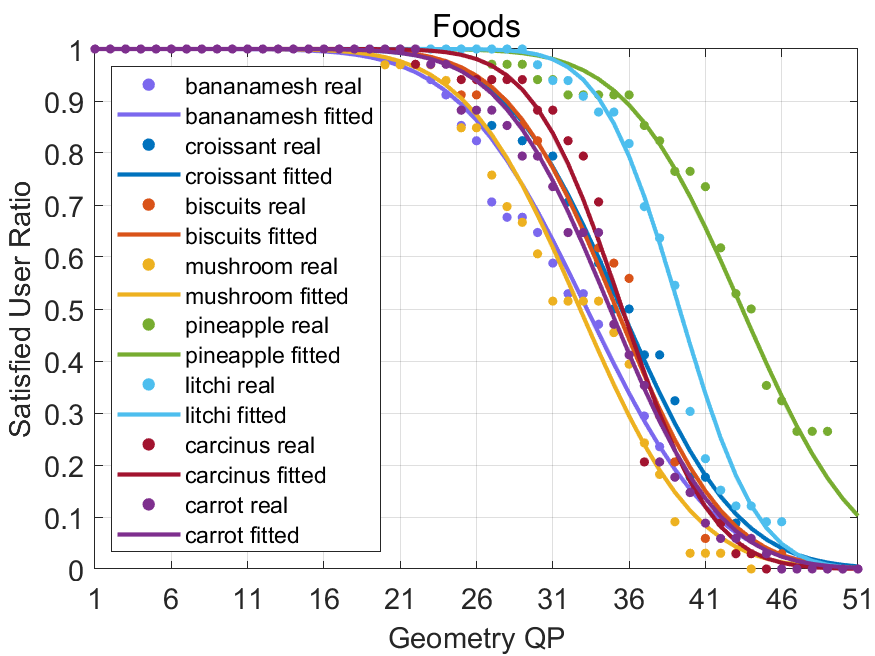}
\includegraphics[width=0.3\textwidth]{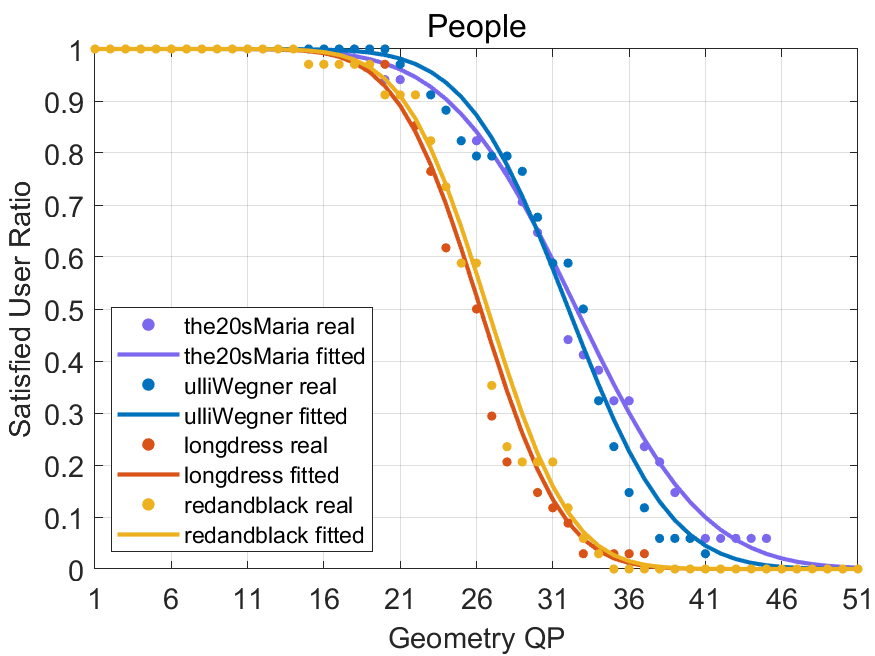}
\includegraphics[width=0.3\textwidth]{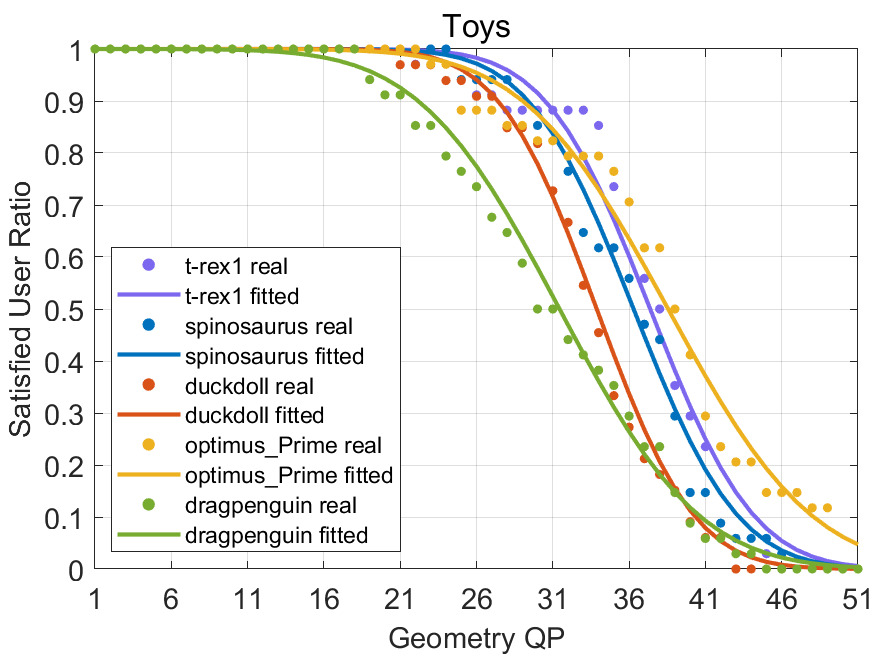}
\includegraphics[width=0.3\textwidth]{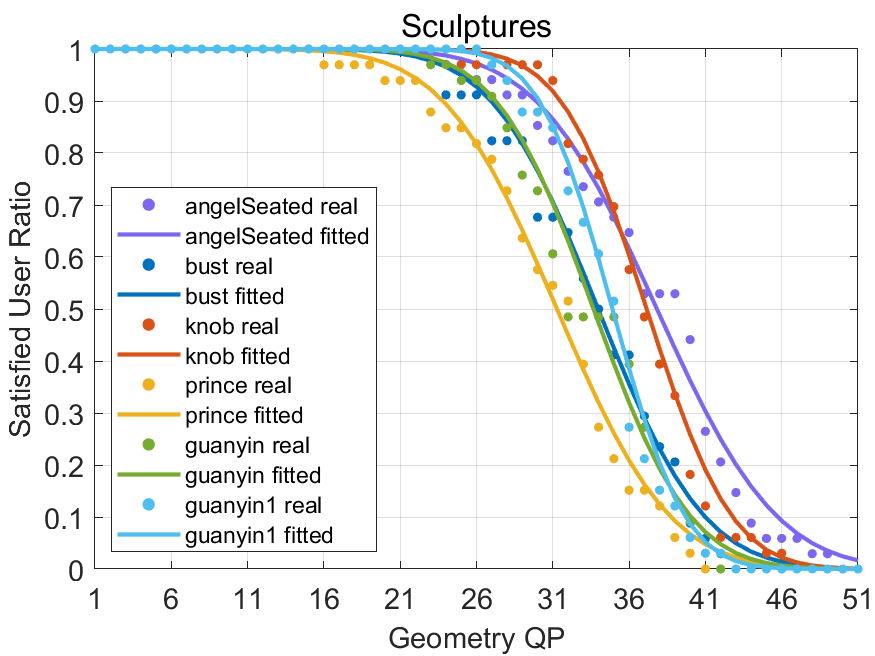}
\includegraphics[width=0.3\textwidth]{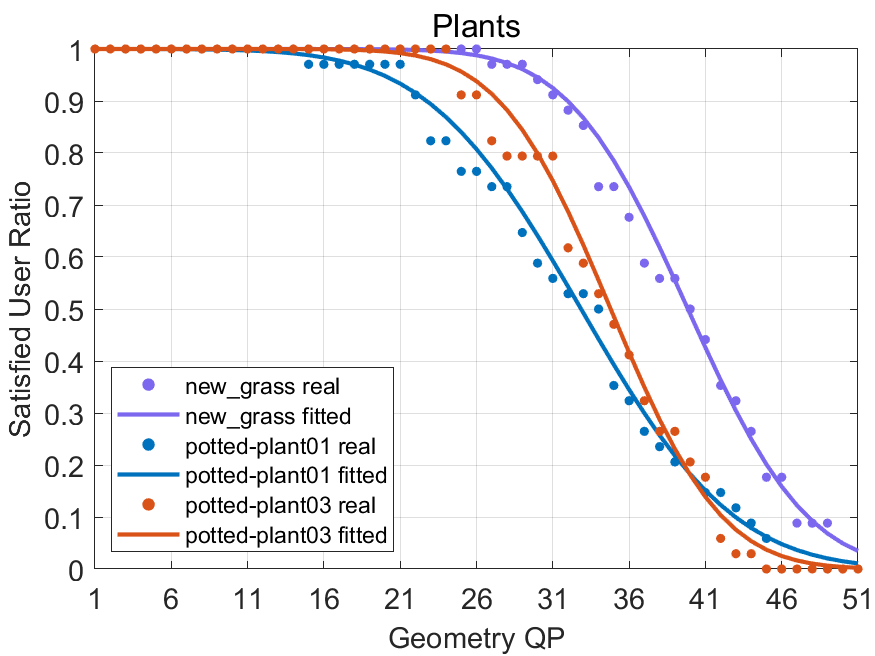}
\includegraphics[width=0.3\textwidth]{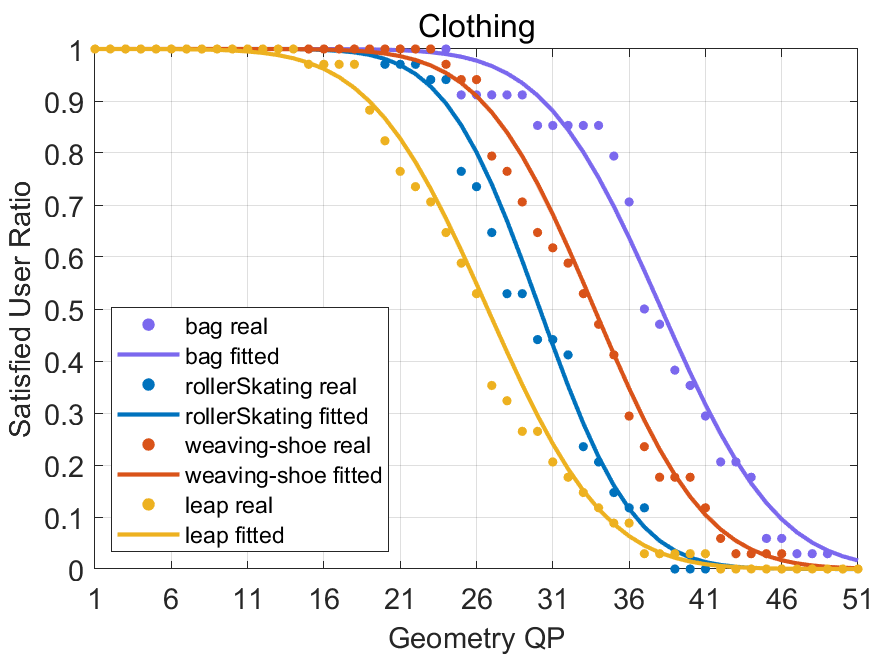}
\centerline{(b)}
\end{minipage}
\caption{SUR as a function of the (a) texture QP, (b) geometry QP. Each figure corresponds to one category of point clouds.} 
\label{fig:SUR}
\vspace{-10pt}
\end{figure*}
Since most references passed the normality test described in Section \ref{Sec:post-processing}, we can assume that the PCJND samples collected for each reference were normally distributed. The Probability Density Function (PDF) of PCJND values obeys normally distribution. We then modeled the SUR curve as the complementary cumulative distribution function of the PDF, and fitted the SUR curves by determining two parameters of its normal distribution, which are mean $\mu$ and standard deviation $\sigma$. Finally, 75\% PCJND can be derived. Figure~\ref{fig:example}
presents the fitted SUR, 75\% and 50\% PCJNDs for point cloud ``bananamesh".
Figure~\ref{fig:SUR} presents the real collected SUR (plotted as points) and fitted SUR curves (plotted as line) for six categories of the point clouds, where (a) is for texture distortion only and (b) is for geometry distortion only. The SUR curves are grouped based on their categories, including Foods, People, Toys, Scuptures, Plants and Clothing. The real SUR points and the fitted SUR curve for one reference are drawn in the same color. SUR curves of different reference point clouds are in different color. We can observe that the SUR curves of most references are well fitted, where each curve reduces from 1 to 0 when the QP increases from 1 to 51. Moreover, the SUR curves vary with contents of reference point cloud in either texture or geometry distortion. For example, the SUR is steeper for ``UliWegner" than ``the20sMaria" in People category. The situations may be different in texture only and geometry distortion only. In addition, the SUR curves of most categories are steeper for texture distorted point clouds than those of geometry distorted point clouds.

\begin{figure}[h]
\centering
\begin{minipage}{0.24\textwidth}
\includegraphics[width=1\textwidth]{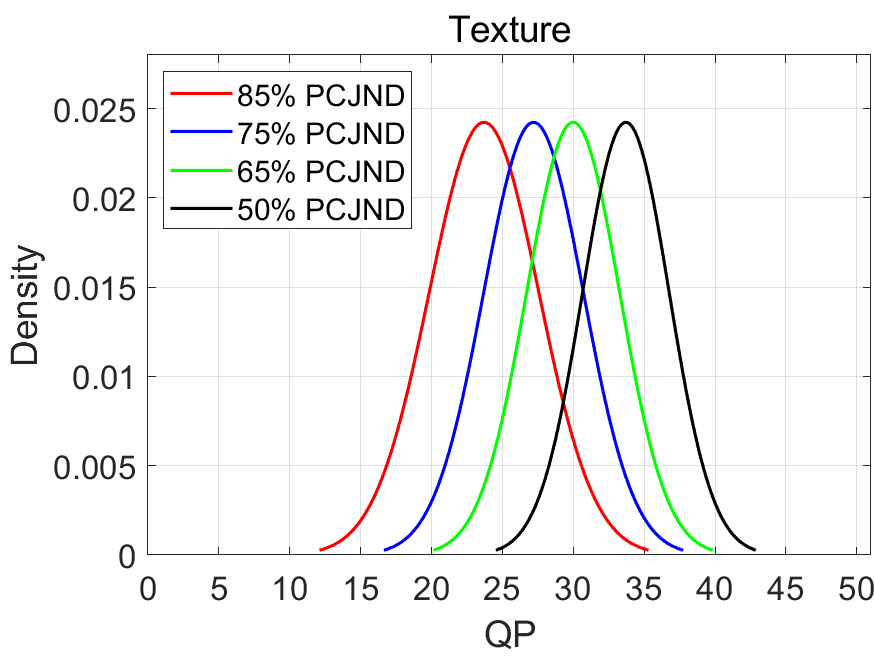}
\centerline{(a)}
\end{minipage}
\begin{minipage}{0.24\textwidth}
\includegraphics[width=1\textwidth]{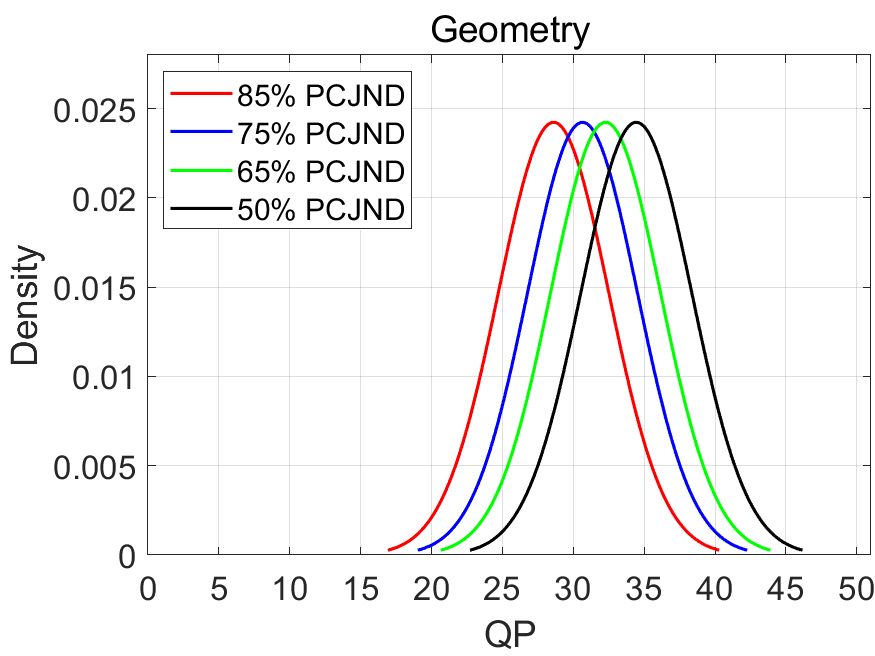}
\centerline{(b)}
\end{minipage}
\caption{Fitting models of the histogram of 
 texture and geometry PCJNDs samples.(a)Texture PCJND, (b)Geometry PCJND.}
\label{fig:PCJNDdistribution}
\end{figure}

\begin{table*}[h]
\centering
\caption{Texture and geometry PCJND in terms of coding QP, standard derivation and PSNR.}\label{tab:jndfortest}
\resizebox{0.96\textwidth}{!}{
\begin{tabular}{ccccccrcccccrc}
\hline
\multirow{2}{*}{Category} & \multirow{2}{*}{Reference} & \multicolumn{6}{c}{texture PCJND } & \multicolumn{6}{c}{geometry PCJND }\\
& & 85\% & 75\% & 65\% & 50\% & STD & PSNR[dB] & 85\% & 75\% & 65\% & 50\% & STD & D1(dB)\\
\hline

\multirow{8}{*}{Foods}
& bananamesh & 21.06 & 24.39 & 27.05 & 30.60 & 9.20 & 21.75 & 26.38 & 28.78 & 30.69 & 33.24 & 6.61 & 59.80\\ 
& croissant & 19.71 & 23.18 & 25.96 & 29.67 & 9.61 & 20.72 & 29.28 & 31.45 & 33.19 & 35.50 & 6.00 & 66.57\\ 
& biscuits & 16.23 & 20.86 & 24.57 & 29.50 & 12.81 & 18.58 & 29.35 & 31.38 & 33.01 & 35.18 & 5.62 & 66.91\\ 
& mushroom & 21.60 & 25.56 & 28.72 & 32.93 & 10.94 & 12.57 & 26.64 & 28.78 & 30.50 & 32.79 & 5.94 & 65.00\\ 
& pineapple & 22.16 & 25.62 & 28.38 & 32.07 & 9.56 & 21.04 & 37.21 & 39.38 & 41.11 & 43.41 & 5.98 & 26.32\\ 
& litchi & 24.03 & 26.87 & 29.15 & 32.17 & 7.86 & 19.79 & 35.12 & 36.58 & 37.75 & 39.30 & 4.03 & 66.49\\ 
& carcinus & 18.96 & 23.28 & 26.73 & 31.33 & 11.93 & 14.17 & 30.77 & 32.44 & 33.78 & 35.56 & 4.62 & 68.55\\ 
& carrot & 25.47 & 29.18 & 32.15 & 36.10 & 10.26 & 17.02 & 28.89 & 30.94 & 32.58 & 34.76 & 5.67 & 66.14\\
\hline

\multirow{4}{*}{People}
& the20sMaria & 17.22 & 21.57 & 25.04 & 29.67 & 12.01 & 19.62 & 25.75 & 28.13 & 30.03 & 32.56 & 6.57 & 63.45\\ 
& ulliWegner & 25.02 & 27.07 & 28.71 & 30.90 & 5.67 & 15.98 & 26.55 & 28.46 & 29.99 & 32.03 & 5.29 & 65.23\\ 
& longdress & 25.18 & 28.02 & 30.28 & 33.30 & 7.83 & 15.16 & 21.84 & 23.38 & 24.62 & 26.26 & 4.27 & 65.80\\ 
& redandblack & 22.13 & 25.82 & 28.77 & 32.70 & 10.20 & 14.11 & 22.31 & 23.85 & 25.09 & 26.74 & 4.27 & 66.08\\ 
\hline

\multirow{5}{*}{Toys}
& t-rex1 & 28.65 & 31.93 & 34.55 & 38.03 & 9.05 & 14.55 & 31.82 & 33.76 & 35.31 & 37.38 & 5.37 & 65.97\\ 
& spinosaurus & 26.53 & 29.59 & 32.04 & 35.30 & 8.46 & 15.06 & 30.71 & 32.66 & 34.22 & 36.29 & 5.39 & 64.47\\ 
& duckdoll & 26.66 & 29.98 & 32.63 & 36.17 & 9.18 & 10.62 & 28.65 & 30.48 & 31.94 & 33.88 & 5.04 & 66.48\\ 
& optimus\_Prime & 29.70 & 32.93 & 35.50 & 38.93 & 8.91 & 15.33 & 30.85 & 33.54 & 35.69 & 38.56 & 7.44 & 62.37\\ 
& dragpenguin & 30.21 & 33.62 & 36.34 & 39.97 & 9.42 & 15.04 & 23.95 & 26.57 & 28.66 & 31.44 & 7.22 & 66.87\\ 
\hline

\multirow{6}{*}{Sculptures}
& angelSeated & 17.41 & 21.99 & 25.65 & 30.53 & 12.66 & 16.95 & 31.42 & 33.66 & 35.44 & 37.82 & 6.18 & 65.84\\ 
& bust & 26.53 & 29.78 & 32.37 & 35.83 & 8.97 & 16.10 & 28.30 & 30.28 & 31.86 & 33.97 & 5.47 & 69.64\\ 
& knob & 23.18 & 26.71 & 29.54 & 33.30 & 9.77 & 14.23 & 32.60 & 34.19 & 35.46 & 37.15 & 4.40 & 66.79\\ 
& prince & 24.33 & 28.31 & 31.49 & 35.73 & 11.00 & 9.22 & 25.23 & 27.34 & 29.03 & 31.27 & 5.83 & 67.07\\ 
& guanyin & 21.41 & 24.99 & 27.85 & 31.67 & 9.90 & 15.44 & 28.49 & 30.30 & 31.74 & 33.67 & 4.99 & 67.59\\ 
& guanyin1 & 23.77 & 27.33 & 30.18 & 33.97 & 9.84 & 16.82 & 31.04 & 32.39 & 33.47 & 34.91 & 3.74 & 67.83\\ 
\hline

\multirow{3}{*}{Plants}
& new\_grass & 24.62 & 28.27 & 31.18 & 35.07 & 10.08 & 16.15 & 33.48 & 35.71 & 37.49 & 39.85 & 6.15 & 60.59\\ 
& potted-plant01 & 27.83 & 30.71 & 33.01 & 36.07 & 7.95 & 12.65 & 24.68 & 27.53 & 29.81 & 32.85 & 7.89 & 67.92\\ 
& potted-plant03 & 21.69 & 25.98 & 29.40 & 33.97 & 11.84 & 14.43 & 28.87 & 30.94 & 32.59 & 34.79 & 5.71 & 67.44\\ 

\hline
\multirow{4}{*}{Clothing}
& bag & 25.61 & 29.38 & 32.39 & 36.40 & 10.41 & 13.21 & 31.86 & 34.05 & 35.79 & 38.12 & 6.03 & 64.24\\ 
& rollerSkating & 19.15 & 23.32 & 26.66 & 31.10 & 11.53 & 7.02 & 25.08 & 26.85 & 28.26 & 30.15 & 4.89 & 66.49\\ 
& weaving-shoe & 25.38 & 28.72 & 31.38 & 34.93 & 9.22 & 14.47 & 27.75 & 29.84 & 31.51 & 33.74 & 5.78 & 65.42\\ 
& leap & 24.27 & 26.78 & 28.79 & 31.47 & 6.95 & 16.20 & 20.45 & 22.65 & 24.40 & 26.74 & 6.06 & 66.89\\ 
\hline

\multirow{4}{*}{Others}
& stone & 17.17 & 20.81 & 23.72 & 27.60 & 10.06 & 18.93 & 33.04 & 35.30 & 37.10 & 39.50 & 6.23 & 68.75\\ 
& coffee\_cup & 27.08 & 30.12 & 32.55 & 35.79 & 8.41 & 14.29 & 26.14 & 28.23 & 29.90 & 32.12 & 5.77 & 66.79\\ 
& house & 24.76 & 27.89 & 30.39 & 33.72 & 8.65 & 18.89 & 33.16 & 35.18 & 36.79 & 38.94 & 5.58 & 66.45\\ 
& ship & 31.21 & 34.16 & 36.51 & 39.66 & 8.15 & 15.80 & 25.06 & 27.01 & 28.57 & 30.65 & 5.39 & 65.99\\ 
\hline
& Mean & 23.70 & 27.20 & 29.99 & 33.71 & 9.65 & 15.64 & 28.61 & 30.65 & 32.28 & 34.44 & 5.63 & 64.83 \\ 
\hline
\end{tabular}
}
\label{Table:PCJND}
\vspace{-10pt}
\end{table*}

\paragraph{Comparison between texture PCJND and geometry PCJND}
The JND is the smallest QP that meets the SUR threshold, $T$~\cite{MCL-JCV}. In this paper, four distinct SUR thresholds, $T\in$ 85\%, 75\%, 65\%, 50\%, were considered in deriving the PCJND. As presented in Table~\ref{Table:PCJND}, we obtained 85\%, 75\%, 65\%, and 50\% texture and geometry PCJNDs by fitting the SUR curves and calculated standard deviation (STD) based on the PCJND samples. In addition, we estimated the quality at 75\% PCJND point using MSE, PSNR (p2point), which is a point-based objective PC quality metric proposed in MPEG point cloud compression software. Firstly, while given stricter threshold $T$, i.e., 85\%, smaller PCJND will be obtained. Secondly, we found that the texture PCJND values of most references are smaller than their geometry PCJND values. Specifically, there are 28 references (82.35\%) at 85\% PCJND, 26 references (76.47\%) at 75\% PCJND, 24 references (70.59\%) at 65\% PCJND, and 18 references (52.94\%) at 50\% PCJND. 
Moreover, the mean texture PCJND values are all smaller than the mean geometry PCJND values at four percentages. Thus, if QP is used to measure distortion, the texture PCJND threshold is smaller than the geometry PCJND threshold. Figure~\ref{fig:PCJNDdistribution} presents the fitting models of the histograms of texture and geometry PCJNDs samples at four percentages (85\%, 75\%, 65\%, and 50\%). We found that the texture PCJND are more concentrated than geometry PCJND. This is consistent with the changing trends in the SUR curves.

To further compare texture PCJND and geometry PCJND, we conducted a paired t-test~\cite{Ttest} and calculated the effect size using Cohen’s \emph{d}~\cite{effectSize} for the PCJNDs at four SUR thresholds $T$. We used a paired t-test to test the hypothesis that the two sample groups come from distributions with equal means. As presented in Table~\ref{tab:t-test}, the t-test results between texture and geometry PCJNDs at 50\%, 65\%, 75\%, and 85\% were 0, 1, 1, and 1. Value ``0'' indicates that the null hypothesis cannot be rejected at the 5\% significance level while value ``1'' indicates that the null hypothesis can be rejected. Cohen’s \emph{d} measures the difference between the mean of texture and geometry PCJNDs at four SUR thresholds. It reports the size of the mean difference compared to the variability of the data. As presented in Table~\ref{tab:t-test}, the effect size results between texture and geometry PCJNDs at 50\%, 65\%, 75\%, and 85\% were 0.1448, 0.4301, 0.6219, and 0.8326. These results were all smaller than one, indicating that the mean difference was smaller compared to the variability of the data. However, as the SUR threshold increased, the effect size becomes larger, meaning that the difference between texture and geometry PCJND was greater.

\begin{table}[h]\caption{Paired t-test and Cohen’s d between texture and geometry PCJNDs.}\label{tab:t-test}
\begin{center}
\begin{tabular}{cccccc}
\hline
\multicolumn{2}{l}{Paired t-test/}& \multicolumn{4}{c}{Geometry PCJND} \\
\multicolumn{2}{l}{Cohen's d} & 50\% & 65\% & 75\% & 85\% \\
\hline
\multirow{4}{*}{Texture PCJND} & 50\% & 0/0.1448 & - & - &- \\
\cline{2-6}
& 65\% & - & 1/0.4301& - & - \\
\cline{2-6}
& 75\% & - & - & 1/0.6219 & - \\
\cline{2-6}
& 85\% & - & - & - & 1/0.8326\\
\hline
\end{tabular}
\end{center}
\vspace{-10pt}
\end{table}

\begin{figure*}[h]
\begin{minipage}{0.24\textwidth}
\includegraphics[width=1\textwidth]{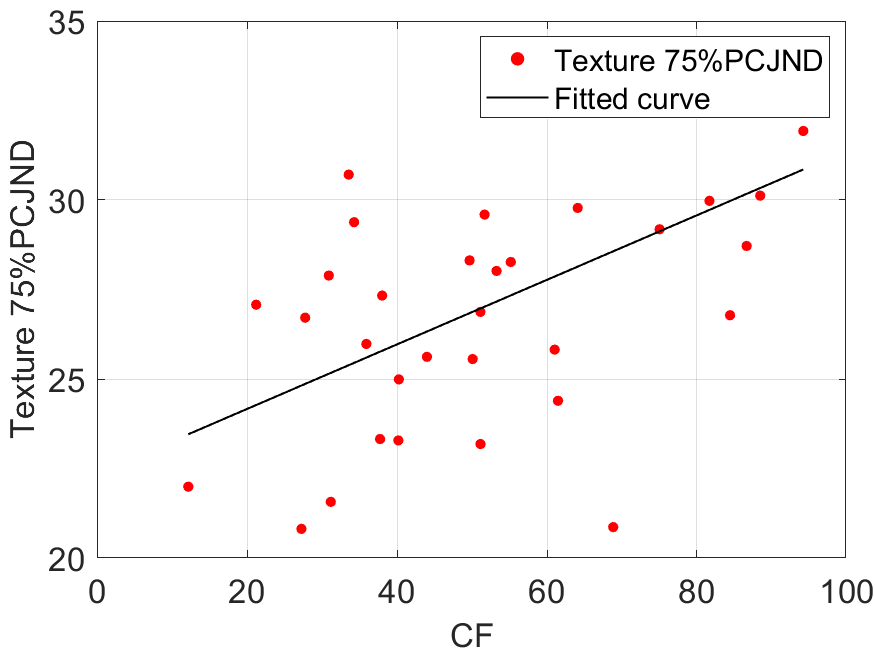}
\centerline{(a)}
\end{minipage}
\begin{minipage}{0.24\textwidth}
\includegraphics[width=1\textwidth]{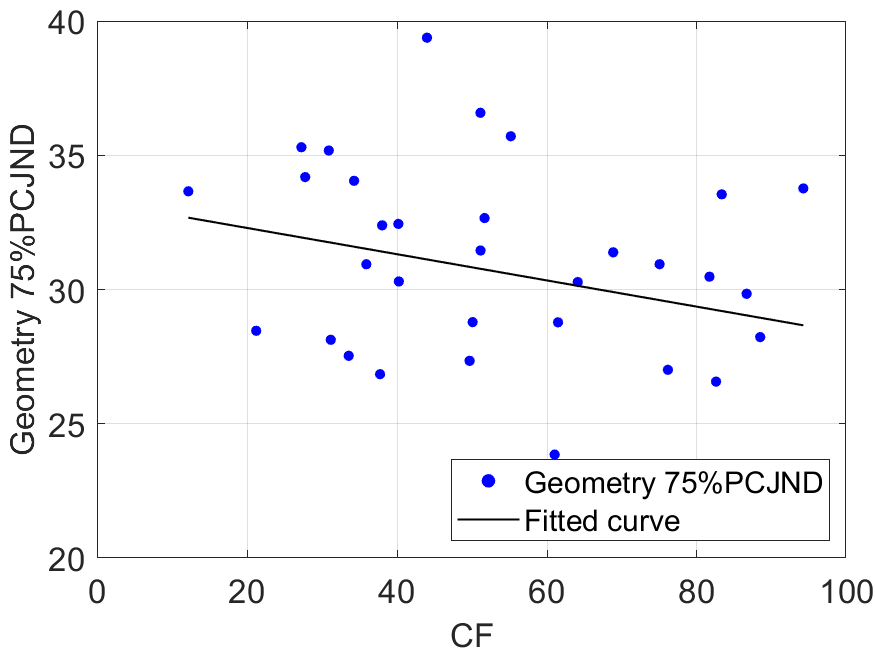}
\centerline{(b)}
\end{minipage}
\begin{minipage}{0.24\textwidth}
\includegraphics[width=1\textwidth]{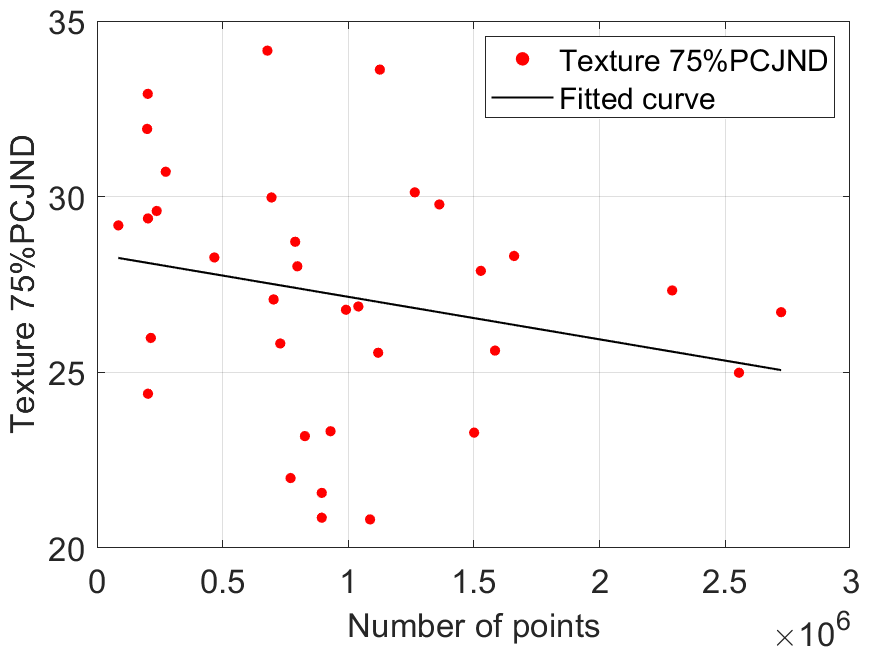}
\centerline{(c)}
\end{minipage}
\begin{minipage}{0.24\textwidth}
\includegraphics[width=1\textwidth]{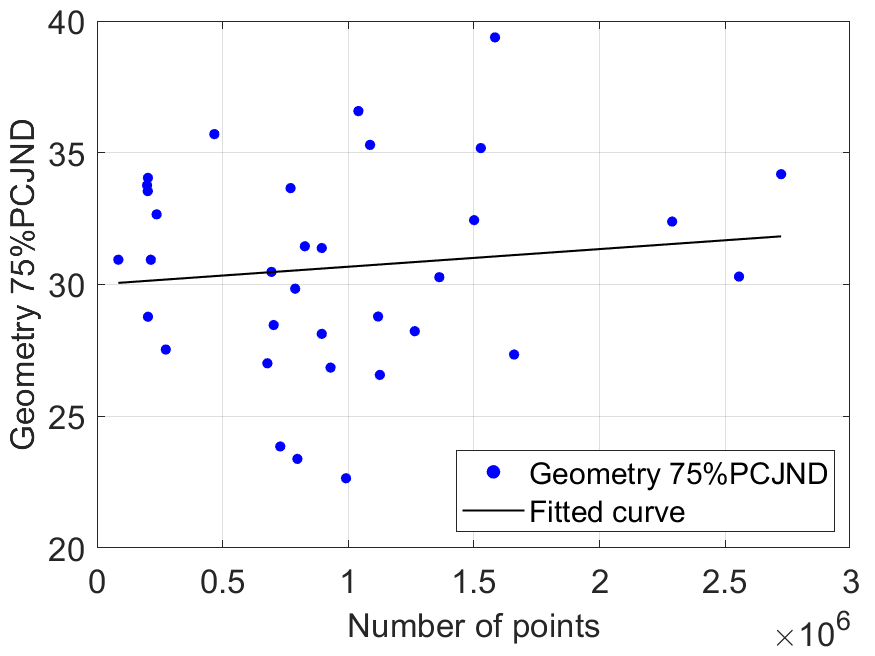}
\centerline{(d)}
\end{minipage}
\caption{Correlation between colorfulness, number of points and PCJND. (a) colorfulness vs. texture PCJND. (b) colorfulness vs. geometry PCJND. (c) Number of points vs. texture PCJND. (d) Number of points vs. geometry PCJND.}\label{fig:CFVS}
\end{figure*}

\paragraph{Correlation between colorfulness, Number of points and PCJND}
To explore the factors that effect the texture and geometry PCJND, we calculated the correlation between colorfulness of texture, number of points and PCJND, respectively. As illustrated in Section~\ref{Sec:referenceCreation}, colorfulness index was calculated for each reference 3D point cloud. As presented in Table~\ref{tab:dataset}, the number of points vary among point clouds. Figure~\ref{fig:CFVS} presents the scatter points and the linear fitted curves between colorfulness, number of points and PCJND. We found that the correlation between CF and texture PCJND is 0.5655, which indicates that the texture PCJND generally increases as the colorfulness increases. Moreover, it is found there is no obvious correlation between colorfulness and geometry PCJND as the correlation index is -0.2784. This may be because colorfulness only estimates perceptual variety and intensity of color, which is generally independent to the geometry.
We also found that there is no obvious correlation between the number of points of source point cloud and texture PCJND, between the number of points and geometry PCJND, where the correlation scores are -0.2279 and 0.1145, respectively. For the same reference point cloud, the more points with texture properties, the richer details may be shown. Geometric compression may result in a loss of points. However, the number of points of source point cloud is used as the initial reference, whose perceptual quality is also related with visual distance, object size, point loss rate from geometry distortion and so on. In other words, the number of points may have less influence on visual perception between references than other factors, such as texture masking effects and others.

\begin{table}[h]\caption{Comparison of the mean and standard deviation of the JND samples.}\label{tab:MeanVSstd}
\begin{center}
\resizebox{0.5\textwidth}{!}{
\begin{tabular}{ccccccc}
\hline
\multirow{2}{*}{datasets} & \multicolumn{3}{c}{Mean} & \multicolumn{3}{c}{Standard deviation} \\
 & min & max & range & min & max & range\\
\hline
VideoSet~\cite{VideoSet} & 19.69 & 36.63 & 16.94 & 2.36 & 9.98 & 7.62\\
Shen~\textit{et al.}~\cite{shen2020jnd} & 23.25 & 40.35 & 17.10 & 3.28 & 9.15 & 5.87\\
SIAT-JSSI~\cite{SIAT-JSSI} & 28.93 & 36.04 & 7.11 & 3.87 & 8.42 & 4.55\\
SIAT-JASI~\cite{SIAT-JASI} & 30.82 & 39.19 & 8.37 & 4.02 & 12.53 & 8.51\\
MCL-JCV~\cite{MCL-JCV} & 21.14 & 31.12 & 9.98 & 3.82 & 7.06 & 3.24\\
PC-JND (proposed) & 26.26 & 43.41 & 14.15 & 3.74 & 12.81 & 9.07 \\
\hline
\end{tabular}
}
\end{center}
\vspace{-10pt}
\end{table}

\paragraph{Limitations}
During pre-training, subjects were instructed to judge whether they noticed a visual difference based on their initial impression within ten seconds. If detecting a difference was difficult, they could choose ``No". However, one potential problem is that participants may have biases in reporting differences due to different visual acuity, preference and reporting criteria. We identified and removed outlier subjects and samples to ensure the reliability of the remaining results. The final PCJND thresholds are statistical quantities. In the future, we will explore methods to overcome these biases.

\section{Conclusion}\label{Sec:concluision}
We conducted a subjective test to study the Point Cloud Just Noticeable Difference (PCJND) in a 6DoF virtual reality environment. To the best of our knowledge, this is the first work of its kind. The subjective test focused on texture and geometry distortion, respectively, in Video based Point Cloud Compression (V-PCC). We developed a platform for our subjective study. We found that the texture PCJND values of most references were smaller than their geometry PCJND values. We also found a correlation between Colorfulness (CF) and texture PCJND. However, we did not find a clear correlation between CF and geometry PCJND, or between the number of points and either texture or geometry PCJND. These PCJND and labels can be used in visual quality assessment and point cloud compression to minimize the bitrate. Moreover, we constructed a PC-JND dataset, which consists of 34 high-quality reference point clouds with their distorted versions encoded using V-PCC. This dataset can serve as a benchmark for developing accurate objective models to predict the PCJND.

\ifCLASSOPTIONcaptionsoff
  \newpage
\fi



\bibliographystyle{IEEEtran}
%
\bibliography{sample-base}

\begin{thebibliography}{10}
\providecommand{\url}[1]{#1}
\csname url@samestyle\endcsname
\providecommand{\newblock}{\relax}
\providecommand{\bibinfo}[2]{#2}
\providecommand{\BIBentrySTDinterwordspacing}{\spaceskip=0pt\relax}
\providecommand{\BIBentryALTinterwordstretchfactor}{4}
\providecommand{\BIBentryALTinterwordspacing}{\spaceskip=\fontdimen2\font plus
\BIBentryALTinterwordstretchfactor\fontdimen3\font minus
  \fontdimen4\font\relax}
\providecommand{\BIBforeignlanguage}[2]{{%
\expandafter\ifx\csname l@#1\endcsname\relax
\typeout{** WARNING: IEEEtran.bst: No hyphenation pattern has been}%
\typeout{** loaded for the language `#1'. Using the pattern for}%
\typeout{** the default language instead.}%
\else
\language=\csname l@#1\endcsname
\fi
#2}}
\providecommand{\BIBdecl}{\relax}
\BIBdecl

\bibitem{wu2019survey}
J.~Wu, G.~Shi, and W.~Lin, ``Survey of visual just noticeable difference
  estimation,'' \emph{Frontiers of Computer Science}, vol.~13, no.~1, pp.
  4--15, Feb. 2019.

\bibitem{LinWeisi}
W.~Lin and G.~Ghinea, ``Progress and opportunities in modelling just-noticeable
  difference ({JND}) for multimedia,'' \emph{IEEE Transactions on Multimedia},
  vol.~24, pp. 3706--3721, Aug. 2022.

\bibitem{Zhang2021Survey}
Y.~Zhang, L.~Zhu, G.~Jiang, S.~T.~W. Kwong, and C.-C.~J. Kuo, ``A survey on
  perceptually optimized video coding,'' \emph{ACM Computing Surveys}, vol.~55,
  no.~12, pp. 1--37, Mar. 2023.

\bibitem{JNDIQA}
S.~Seo, S.~Ki, and M.~Kim, ``A novel just-noticeable-difference-based
  saliency-channel attention residual network for full-reference image quality
  predictions,'' \emph{IEEE Transactions on Circuits and Systems for Video
  Technology}, vol.~31, no.~7, pp. 2602--2616, Oct. 2021.

\bibitem{JNDcoding}
X.~Zhang, S.~Wang, K.~Gu, W.~Lin, S.~Ma, and W.~Gao, ``Just-noticeable
  difference-based perceptual optimization for jpeg compression,'' \emph{IEEE
  Signal Processing Letters}, vol.~24, no.~1, pp. 96--100, Jan. 2017.

\bibitem{SURcoding}
Z.~Yang, W.~Gao, G.~Li, and Y.~Yan, ``Sur-driven video coding rate control for
  jointly optimizing perceptual quality and buffer control,'' \emph{IEEE
  Transactions on Image Processing}, vol.~32, pp. 5451--5464, Sep. 2023.

\bibitem{JNDenhancement}
C.-H. Lee, P.-Y. Lin, L.-H. Chen, and W.-K. Wang, ``Image enhancement approach
  using the just-noticeable-difference model of the human visual system,''
  \emph{Journal of Electronic Imaging}, vol.~21, no.~3, pp. 1--14, July 2012.

\bibitem{watermarking}
H.~Fang, Z.~Jia, H.~Zhou, Z.~Ma, and W.~Zhang, ``Encoded feature enhancement in
  watermarking network for distortion in real scenes,'' \emph{IEEE Transactions
  on Multimedia}, vol.~25, pp. 2648--2660, Feb. 2023.

\bibitem{WuEnhancedJND}
J.~Wu, L.~Li, W.~Dong, G.~Shi, W.~Lin, and C.-C.~J. Kuo, ``Enhanced just
  noticeable difference model for images with pattern complexity,'' \emph{IEEE
  Transactions on Image Processing}, vol.~26, no.~6, pp. 2682--2693, June 2017.

\bibitem{ZENG}
Z.~Zeng, H.~Zeng, J.~Chen, J.~Zhu, Y.~Zhang, and K.-K. Ma, ``Visual attention
  guided pixel-wise just noticeable difference model,'' \emph{IEEE Access},
  vol.~7, pp. 132\,111--132\,119, Sep. 2019.

\bibitem{wang2020improved}
H.~Wang, L.~Yu, H.~Yin, T.~Li, and S.~Wang, ``An improved {DCT}-based {JND}
  estimation model considering multiple masking effects,'' \emph{Journal of
  Visual Communication and Image Representation}, vol.~71, p. 102850, Aug.
  2020.

\bibitem{MCL-JCI}
L.~Jin, J.~Y. Lin, S.~Hu, H.~Wang, P.~Wang, I.~Katsavounidis, A.~Aaron, and
  C.-C.~J. Kuo, ``Statistical study on perceived jpeg image quality via
  {MCL-JCI} dataset construction and analysis,'' \emph{Electronic Imaging},
  vol. 2016, no.~13, pp. 1--9, 2016.

\bibitem{liuPWJND}
H.~Liu, Y.~Zhang, H.~Zhang, C.~Fan, S.~Kwong, C.-C.~J. Kuo, and X.~Fan, ``Deep
  learning-based picture-wise just noticeable distortion prediction model for
  image compression,'' \emph{IEEE Transactions on Image Processing}, vol.~29,
  pp. 641--656, Aug. 2019.

\bibitem{zhangVWJND}
Y.~Zhang, H.~Liu, Y.~Yang, X.~Fan, S.~Kwong, and C.-C.~J. Kuo, ``Deep learning
  based just noticeable difference and perceptual quality prediction models for
  compressed video,'' \emph{IEEE Transactions on Circuits and Systems for Video
  Technology}, vol.~32, no.~3, pp. 1197--1212, Apr. 2021.

\bibitem{fan2021multimedia}
C.~Fan, Y.~Zhang, R.~Hamzaoui, D.~Ziou, and Q.~Jiang, ``Learning-based
  satisfied user ratio prediction for symmetrically and asymmetrically
  compressed stereoscopic images,'' \emph{IEEE MultiMedia}, vol.~28, no.~3, pp.
  8--20, Feb 2021.

\bibitem{fanSURnet}
C.~Fan, H.~Lin, V.~Hosu, Y.~Zhang, Q.~Jiang, R.~Hamzaoui, and D.~Saupe,
  ``{SUR}-{N}et: Predicting the satisfied user ratio curve for image
  compression with deep learning,'' in \emph{eleventh International Conference
  on Quality of Multimedia Experience}, Berlin, Germany, June 2019, pp. 1--6.

\bibitem{RecommendationITUR}
ITU-R, ``Methodologies for the subjective assessment of the quality of
  television images,'' \emph{Recommendation ITU-R BT.500-15}, 2023.

\bibitem{VPCC}
\BIBentryALTinterwordspacing
D.~B. Graziosi, O.~Nakagami, S.~Kuma, A.~Zaghetto, S.~Teruhiko, and A.~J.
  Tabatabai, ``An overview of ongoing point cloud compression standardization
  activities: video-based (v-pcc) and geometry-based (g-pcc),'' \emph{APSIPA
  Transactions on Signal and Information Processing}, vol.~9, 2020. [Online].
  Available: \url{https://api.semanticscholar.org/CorpusID:216453743}
\BIBentrySTDinterwordspacing

\bibitem{shen2020jnd}
X.~Shen, Z.~Ni, W.~Yang, X.~Zhang, S.~Wang, and S.~Kwong, ``A {JND} dataset
  based on {VVC} compressed images,'' in \emph{2020 IEEE International
  Conference on Multimedia \& Expo Workshops}, London, UK, July 2020, pp. 1--6.

\bibitem{KonJND-1k}
H.~Lin, G.~Chen, M.~Jenadeleh, V.~Hosu, U.-D. Reips, R.~Hamzaoui, and D.~Saupe,
  ``Large-scale crowdsourced subjective assessment of picturewise just
  noticeable difference,'' \emph{IEEE Transactions on Circuits and Systems for
  Video Technology}, vol.~32, no.~9, pp. 5859--5873, Mar. 2022.

\bibitem{MCL-JCV}
H.~Wang, W.~Gan, S.~Hu, J.~Y. Lin, L.~Jin, L.~Song, P.~Wang, I.~Katsavounidis,
  A.~Aaron, and C.-C.~J. Kuo, ``{MCL-JCV}: a {JND}-based {H}. 264/{AVC} video
  quality assessment dataset,'' in \emph{IEEE international Conference on Image
  Processing}, Phoenix, AZ, USA, Sep. 2016, pp. 1509--1513.

\bibitem{VideoSet}
H.~Wang, I.~Katsavounidis, J.~Zhou, J.~Park, S.~Lei, X.~Zhou, M.-O. Pun,
  X.~Jin, R.~Wang, X.~Wang \emph{et~al.}, ``{V}ideo{S}et: A large-scale
  compressed video quality dataset based on jnd measurement,'' \emph{Journal of
  Visual Communication and Image Representation}, vol.~46, pp. 292--302, 2017.

\bibitem{HD-VJND}
J.~Zhu, A.-F. Perrin, and P.~L. Callet, ``Subjective test methodology
  optimization and prediction framework for just noticeable difference and
  satisfied user ratio for compressed {HD} video,'' in \emph{Picture Coding
  Symposium}, San Jose, CA, USA, Jan. 2022, pp. 313--317.

\bibitem{SIAT-JSSI}
C.~Fan, Y.~Zhang, R.~Hamzaoui, and Q.~Jiang, ``Interactive subjective study on
  picture-level just noticeable difference of compressed stereoscopic images,''
  in \emph{IEEE International Conference on Acoustics, Speech and Signal
  Processing}, Brighton, UK, May 2019, pp. 8548--8552.

\bibitem{SIAT-JASI}
C.~Fan, Y.~Zhang, H.~Zhang, R.~Hamzaoui, and Q.~Jiang, ``Picture-level just
  noticeable difference for symmetrically and asymmetrically compressed
  stereoscopic images: Subjective quality assessment study and datasets,''
  \emph{Journal of Visual Communication and Image Representation}, vol.~62, pp.
  140--151, July 2019.

\bibitem{JND-Pano}
X.~Liu, Z.~Chen, X.~Wang, J.~Jiang, and S.~Kowng, ``{JND}-{P}ano: Database for
  just noticeable difference of {JPEG} compressed panoramic images,'' in
  \emph{19th Pacific-Rim Conference on Multimedia}, Hefei, China, Sep. 2018,
  pp. 458--468.

\bibitem{SI-CF}
\BIBentryALTinterwordspacing
D.~Hasler and S.~Susstrunk, ``Measuring colourfulness in natural images,'' in
  \emph{Human Vision and Electronic Imaging VIII}, vol. 5007, Santa Clara, CA,
  United states, Jan. 2003, pp. 87--95. [Online]. Available:
  \url{http://dx.doi.org/10.1117/12.477378}
\BIBentrySTDinterwordspacing

\bibitem{VPCCandGPCC}
D.~B. Graziosi, O.~Nakagami, S.~Kuma, A.~Zaghetto, S.~Teruhiko, and
  A.~Tabatabai, ``An overview of ongoing point cloud compression
  standardization activities: video-based ({V-PCC}) and geometry-based
  ({G-PCC}),'' \emph{APSIPA Transactions on Signal and Information Processing},
  vol.~9, pp. 1--17, Apr. 2020.

\bibitem{Perry}
S.~Perry, H.~P. Cong, L.~A. da~Silva~Cruz, J.~Prazeres, M.~Pereira,
  A.~Pinheiro, E.~Dumic, E.~Alexiou, and T.~Ebrahimi, ``Quality evaluation of
  static point clouds encoded using mpeg codecs,'' in \emph{2020 IEEE
  International Conference on Image Processing}, 2020, pp. 3428--3432.

\bibitem{wusubjective}
X.~Wu, Y.~Zhang, C.~Fan, J.~Hou, and S.~Kwong, ``Subjective quality database
  and objective study of compressed point clouds with 6{DoF} head-mounted
  display,'' \emph{IEEE Transactions on Circuits and Systems for Video
  Technology}, vol.~31, no.~12, pp. 4630--4644, July 2021.

\bibitem{guo2016subjective}
J.~Guo, V.~Vidal, I.~Cheng, A.~Basu, A.~Baskurt, and G.~Lavoue, ``Subjective
  and objective visual quality assessment of textured 3d meshes,'' \emph{ACM
  Transactions on Applied Perception}, vol.~14, no.~2, pp. 1--20, Oct. 2016.

\bibitem{comparisonVR}
Y.~Nehm{\'e}, J.-P. Farrugia, F.~Dupont, P.~L. Callet, and G.~Lavou{\'e},
  ``Comparison of subjective methods for quality assessment of 3{D} graphics in
  virtual reality,'' \emph{ACM Transactions on Applied Perception}, vol.~18,
  no.~1, pp. 1--23, Dec. 2020.

\bibitem{grubbs}
F.~E. Grubbs \emph{et~al.}, ``Sample criteria for testing outlying
  observations,'' \emph{The Annals of Mathematical Statistics}, vol.~21, no.~1,
  pp. 27--58, 1950.

\bibitem{JBtest}
\BIBentryALTinterwordspacing
A.~K. Bera and C.~M. Jarque, ``Efficient tests for normality, homoscedasticity
  and serial independence of regression residuals: Monte carlo evidence,''
  \emph{Economics Letters}, vol.~7, no.~4, pp. 313--318, 1981. [Online].
  Available:
  \url{https://www.sciencedirect.com/science/article/pii/0165176581900355}
\BIBentrySTDinterwordspacing

\bibitem{Ttest}
\BIBentryALTinterwordspacing
C.~A. Markowski and E.~P. Markowski, ``Conditions for the effectiveness of a
  preliminary test of variance,'' \emph{The American Statistician}, vol.~44,
  no.~4, pp. 322--326, 1990. [Online]. Available:
  \url{https://www.tandfonline.com/doi/abs/10.1080/00031305.1990.10475752}
\BIBentrySTDinterwordspacing

\bibitem{effectSize}
\BIBentryALTinterwordspacing
L.~Berben, S.~M. Sereika, and S.~Engberg, ``Effect size estimation: Methods and
  examples,'' \emph{International Journal of Nursing Studies}, vol.~49, no.~8,
  pp. 1039--1047, 2012. [Online]. Available:
  \url{https://www.sciencedirect.com/science/article/pii/S0020748912000442}
\BIBentrySTDinterwordspacing

\end{thebibliography}



%








\end{document}